%% file: wiener-proof.tex
\documentclass[openacc]{rsproca_new}

\usepackage{bm}
\usepackage{placeins}
\usepackage{siunitx}

\include{preamble}

\include{macros}
\include{mst-macros}

\graphicspath{{images/}}

\begin{document}

\title{A Proof that Multiple Waves Propagate in Ensemble-Averaged Particulate Materials}
\author{Artur L.\ Gower$^1$, I.\ David Abrahams$^2$, and William J.\ Parnell$^3$}

\address{$^1$Department of Mechanical Engineering, The University of Sheffield, UK, \\
$^2$ Isaac Newton Institute for Mathematical Sciences, 20 Clarkson Road, Cambridge CB3 0EH, UK\\
$^3$ School of Mathematics, University of Manchester, Oxford Road,
Manchester M13 9PL, UK}

\subject{Wave motion, acoustics, mathematical physics}


\corres{Artur L.\ Gower\\
\email{arturgower$@$gmail.com}\\
website: \href{http://arturgower.github.io}{arturgower.github.io} }

\begin{abstract}
  \rev{Effective medium theory aims to describe a complex inhomogeneous material in terms of a few important macroscopic parameters. To characterise wave propagation through an inhomogeneous material, the most crucial parameter is the \emph{effective wavenumber}. For this reason, there are many published studies on how to calculate a single effective wavenumber. Here we present a proof that there \emph{does not} exist a unique effective wavenumber; instead, there are an infinite number of such (complex) wavenumbers. We show that in most parameter regimes only a small number of these effective wavenumbers make a significant contribution to the wave field. However, to accurately calculate the reflection and transmission coefficients, a large number of the (highly attenuating) effective waves is required. For clarity, we present results for scalar (acoustic) waves for a two-dimensional material filled (over a half space) with randomly distributed circular cylindrical inclusions. We calculate the effective medium by ensemble averaging over all possible inhomogeneities. The proof is based on the application of the Wiener-Hopf technique and makes no assumption on the wavelength, particle boundary conditions/size, or volume fraction. This technique provides a simple formula for the reflection coefficient, which can be explicitly evaluated for monopole scatterers. We compare results with an alternative numerical matching method.}
 \end{abstract}

\maketitle

\section{Introduction}

Materials comprising particles or inclusions that are randomly distributed inside a uniform host medium occur frequently in the world around us. They occur as synthetically fabricated media and also in nature. Common examples include composites, emulsions, suspensions, complex gases, and polymers. Understanding how electromagnetic, elastic, or acoustic waves propagate through these materials is necessary in order to characterise the properties of these materials, and also to design new materials that can control wave propagation.

The wave scattered from a particulate material will be influenced by the positions and properties of all particles, which are usually unknown. However, this scattered field, averaged over space or over time, depends only on the average particle properties. Many measurement systems perform averaging over space, if the receivers or incident wavelength are large enough~\cite{pinfield_emergence_2013}, or over time~\cite{mishchenko_multiple_2006}. In most cases, this averaging process is the same as averaging over all possible particle configurations. Such systems are sometimes called ergodic~\cite{mishchenko_multiple_2006,mishchenko_first-principles_2016}. In this paper, we focus on ensemble averaged waves, satisfying the scalar wave equation in two-dimensions, reflecting from, and propagating in, a half-space particulate material. In certain scenarios, such as light scattering, it is easier to measure the average intensity of the wave. However, even in these cases, the ensemble-averaged field is often needed as a first step~\cite{tsang_radiative_1987,tsang_dense_2000}.

One driving principle\edit{, often} used in the literature, is that the ensemble-averaged wave \edit{itself} satisfies a wave equation \edit{with a} single effective wavenumber~\cite{foldy_multiple_1945,lax_multiple_1951,fikioris_multiple_1964}. Reducing a \edit{inhomogeneous} material, with many unknowns, down to one effective wavenumber is attractive as it greatly reduces the complexity of the problem. For this reason many papers have attempted to deduce this unique effective wavenumber from first principles in electromagnetism~\cite{tsang_scattering_1983,tishkovets_scattering_2011,mishchenko_first-principles_2016}, acoustics~\cite{linton_multiple_2005,linton_multiple_2006,martin_multiple_2011,norris_multiple_2011,gower_reflection_2018} and elasticity~\cite{conoir_effective_2010,norris_effective_2012}. See~\cite{gower_multiple_2018} for a short overview of the history of this topic, including typical statistical assumptions employed within the methods, such as hole-correction and the quasi-crystalline approximation, which we also adopt here.

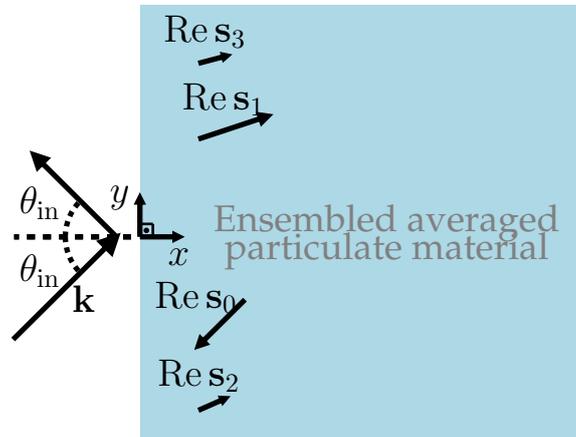
\begin{figure}[htp!]
  \centering
  \input{"images/multiple-waves-diagram.tex"}
  \caption{When an incident plane wave $\ee^{\ii \mathbf k \cdot (x, y)}$, with $\mathbf k = k (\cos \theta_\inc,\sin \theta_\inc)$, encounters an (ensemble-averaged) particulate material, it excites many transmitted plane waves and one reflected plane wave. The transmitted waves are of the form $\ee^{\ii \mathbf s_p \cdot (x,y)}$ with wavenumbers $\mathbf s_p = S_p(\cos \theta_p, \sin \theta_p )$ where both $S_p$ and $\theta_p$ are complex numbers. The larger Im $s_p$, the more quickly the wave attenuates as it propagates into the half-space and the smaller the drawn vector for that wave above. The results shown here represent the effective wavenumbers for parameters~\eqref{eqn:parameters}, which are shown in~\Cref{fig:wavenumbers-example}.  }
  \label{fig:multiple-diagram}
\end{figure}

\edit{The assumption that the ensemble averaged wave field satisfies a wave equation, with an effective wavenumber, has never been fully justified. Here we prove that} there \emph{does not} exist a unique effective wavenumber but instead there are an infinite number of them. Gower et al.\ \cite{gower_multiple_2018} first \edit{showed that there exist many effective wavenumbers, and} provided a technique, the \emph{Matching Method}, to efficiently calculate the effective wave field. In the present paper and \cite{gower_multiple_2018}, we show that for some parameter regimes, at least two effective wavenumbers are needed to obtain accurate results, when compared with numerical simulations. \edit{We also provide examples of how a single effective wave approximation leads to inaccurate results for both transmission and reflection for a halfspace filled with particles, see \Cref{fig:multiple-diagram}.}

Although the \emph{Matching Method} developed in \cite{gower_multiple_2018} gave accurate results, when compared to numerical methods and known asymptotic limits, the limitations of the method were not immediately clear. Here however we illustrate that the Matching Method is robust, because combining many effective wavenumbers is not just a good approximation, it is an analytical solution to the integral equation governing the ensemble averaged wave field. We prove this by employing the Wiener-Hopf technique and then, for clarity, illustrate the solution for particles that scatterer only in their monopole mode. The Wiener-Hopf technique also gives a simple and elegant expression for the reflection coefficient.

The Wiener-Hopf technique is a powerful tool to solve a diverse range of wave scattering problems, see \cite[Chapter 5. Wiener-Hopf Technique]{crighton_modern_1992} and \cite{lawrie_brief_2007,noble_methods_1988} for an introduction. It is especially useful for semi-infinite domains~\cite{martin_one-dimensional_2015,norris_acoustic_1995,haslinger_controlling_2016,tymis_scattering_2014,haslinger_semi-infinite_2017,albani_wave_2011} and boundary value problems of mixed type. In this work, the Wiener-Hopf technique clearly reveals the form of the analytic solution, but to compute the solution would require an analytic factorisation of a matrix-function. To explicitly perform this factorisation is difficult~\cite{abrahams_solution_1997,abrahams_radiation_1996,abrahams_general_1990,rogosin_constructive_2016}. Indeed this is often the hardest aspect of employing the Wiener-Hopf technique, although there exist approximate methods for this purpose~\cite{abrahams_solution_1997,kisil_iterative_2018,abrahams_application_2000,abrahams_scattering_1987}. We do not focus in this article on these analytic factorisations, as there already exists a method to compute the required solution~\cite{gower_multiple_2018}. Instead, the present work acts as proof that the Matching Method~\cite{gower_multiple_2018} \edit{faithfully reproduces the form of the analytic solution.}


\Cref{fig:multiple-diagram} shows the main setup and result of this paper: an incident plane wave excites the half-space $x>0$ filled with ensemble-averaged particles (the blue region), which generates a reflected wave and many effective transmitted waves. The $\vec s_p$ are the transmitted wavevectors, and the smaller the length of the vector, the faster that effective wave attenuates as it propagates further into the material.

The paper begins by summarising the equations that govern ensemble averaged waves in two-dimensions in \cref{sec:ensemble_average}. Following this, in \cref{sec:wiener-hopf} we apply the Wiener-Hopf technique to the governing integral equation and deduce that the solution is a superposition of plane waves, each with a different effective wavenumber. A simple expression for the reflection coefficient is also derived. In \cref{sec:monopole} we specialise the results for particles that scatter only in the monopole mode, which leads to a closed form analytic solution.

\edit{The dispersion relation~\eqref{eqn:dispersion}, derived in \cref{sec:wiener-hopf}, admits an infinite number of solutions, the effective wavenumbers. In \cref{sec:multiple-wavenumbers}, we deduce asymptotic forms for the effective wavenumbers in both a low and high frequency limit.} In \cref{sec:Numerical} we compare numerical results for monopole scatterers, using the Wiener-Hopf technique, with classical methods that assume only one effective wavenumber~\cite{linton_multiple_2005,martin_multiple_2011}, and the Matching Method introduced in~\cite{gower_multiple_2018}. In general, when comparing predicted reflection coefficients, the Wiener-Hopf and Matching Method agree well, whereas the classical single-effective-wavenumber method can disagree by anywhere up to $20\%$. These results are discussed in~\cref{sec:Conclusion} together with anticipated future steps.

\section{Waves in ensemble averaged particles}
\label{sec:ensemble_average}

Consider a region filled with particles or inclusions that are uniformly distributed. The field $u$ is governed by the scalar wave equations:
\begin{align}
  &\nabla^2 u + k^2 u = 0, \quad \text{(in the background material)}, \\
  &\nabla^2 u + k^2_o u = 0, \quad \text{(inside a particle)},
\end{align}
where $k$ and $k_o$ are the real wavenumbers of the background and inclusion materials, respectively. We assume all particles are identical, except for their position and orientation, for simplicity. For a distribution of particles, or multi-species, see~\cite{gower_reflection_2018}.

Our goal is to calculate the ensemble average field $\ensem{u(x,y)}$, that is, the field averaged over all possible particle positions and orientations. For clarity, and ease of exposition, we consider that the particles are equally likely to be located anywhere except that they cannot overlap (this is often called the \emph{hole correction} assumption). We also assume the quasi-crystalline approximation; for details on this, and for further details on deducing the results in this section, see~\cite{linton_multiple_2005,gower_reflection_2018,gower_multiple_2018}.

By splitting the total steady wave field $u(x,y)$ into a sum of the incident wave $u_\inc(x,y)$ and waves scattered by each particle, the $j$th scattered wave being $u_j(x,y)$, we can write:
\begin{equation}
  u(x,y) = u_\inc(x,y) + \sum_{j} u_j(x,y).
\label{eqn:incident_split}
\end{equation}
A simple and useful scenario to consider is when all particles are placed only within the half-space\footnote{The case where particles can be placed anywhere in the plane can lead to ill-defined integrals~\cite{linton_multiple_2005}.} $x > 0$, which are then excited by a plane wave, \rev{with implicit time dependence $\ee^{\ii \omega t}$,} incident from a homogeneous region:
\begin{equation}
  u_\inc (x,y) = \ee^{\ii (\alpha x + \beta y)}, \quad \text{with} \quad (\alpha,\beta) = (k \cos \theta_\inc, k\sin \theta_\inc),
  \label{eqn:incident}
\end{equation}
where we restrict the incident angle $ -\frac{\pi}{2} < \theta_\inc < \frac{\pi}{2}$, as shown in~\Cref{fig:multiple-diagram}, and
 consider a slightly dissipative medium with
\begin{equation}
  \mathrm{Re} \, k >0 \quad \text{and} \quad \mathrm{Im} \, k >0.
  \label{ineq:k}
\end{equation}
This dissipation will facilitate the use of the Wiener-Hopf technique, and after reaching the solution we can take $k$ to be real\footnote{\rev{Assuming Im $k = \epsilon >0$, rather than $\epsilon \geq 0$, will facilitate calculating certain integrals that appear below. However, after reaching a solution, we can take the limit $\epsilon \to 0$ to recover the physically viable solution for Im $k =0$.}}.

To describe the particulate medium we employ the following notation:
\begin{align}
& b = \text{the minimum distance between particle centres},
\\
& \nfrac {} = \text{number of particles per unit area},
\\
& T_n = \text{the coefficients of the particle's T-matrix},
\\
& \phi =  \frac{\pi \nfrac {} b_{}^2}{4} = \text{particle area fraction}.
  \label{eqn:non-dimensional}
\end{align}
Although the area fraction $\phi$, normally called the volume fraction, is a combination of other parameters, it is useful because it is non-dimensional. If we let $a_o$ be the maximum distance from the particle's centre to its boundary, then we can set $b_{} = \gamma a_o$, where $\gamma \geq 2$ so as to avoid two particles overlapping. The volume fraction that does not include the exclusion zone
$\phi'$, as used in \cite[equation (4.7)]{gower_multiple_2018}, is then $\phi' = 4 \phi/\gamma^2$.

The $T_n$ are the coefficients of a diagonal T-matrix~\cite{ganesh_far-field_2010,ganesh_algorithm_2017,mishchenko_light_1991,mishchenko_light_1993,waterman_symmetry_1971}. The T-matrix determines how the particle scatters waves, and so depends on the particle's shape and boundary conditions. A diagonal T-matrix can be used to represent either a radially symmetric particle, or  particles averaged over their orientation, assuming the orientations have a random uniform distribution.

\rev{The results of ensemble averaging~\eqref{eqn:incident_split} from first principals are deduced in a number of references\cite{gower_multiple_2018,gower_reflection_2018} and so deails of this procedure are omitted here for brevity.} To represent the ensemble averaged scattered wave from a particle, whose centre is fixed at $(x_1, y_1)$, we use
\begin{equation}
  \ensem{u_1 (x_1 + X,y_1 + Y)}_{(x_1,y_1)} = \sum_{n = -\infty}^\infty \Ab_n(x_1) \ee^{\ii \beta y_1} \mathrm H^{(1)}_n(k R) \ee^{\ii n\Theta},
\end{equation}
for $R := \sqrt{X^2 + Y^2} > b/2$, so that $(X,Y)$ is on the outside of this particle, with $(R,\Theta)$ being the polar coordinates of $(X,Y)$, $\mathrm H^{(1)}_{n}$ are Hankel functions of the first kind, and $\Ab_n$ is some field we want to determine\footnote{\rev{The factor $\ee^{\ii \beta y_1}$ appears due to the translational invariance of $\ensem{u_1(x_1,y_1)}$ in $y_1$, which is a result of the material being statistically homogeneous, see~\cite{gower_reflection_2018} for details.}}.

By choosing\footnote{We define the reflection coefficient only for $x<-b_{}$, instead of $x < -b/2$, so that we can use $\psi_n$ in the formula for $\mathfrak R$, which will in turn facilitate calculating $\mathfrak R$.
} $x < -b$, which is outside of the region filled with particles, then taking the ensemble average on both sides of~\eqref{eqn:incident_split} results in equation (6.7) of \cite{gower_multiple_2018}, given by:
\begin{equation}
  \ensem{u(x,y)} = u_\inc(x,y) + \mathfrak R \ee^{- \ii \alpha x  + \ii \beta y} \;\;\; \text{for} \;\;\; x < -b_{},
  \label{eqn:AverageWave}
\end{equation}
which is the incident wave plus an effective reflected wave with reflection coefficient:
\begin{equation}
  \mathfrak R = \ee^{ \ii \alpha x} \nfrac {} \sum_{n=-\infty}^\infty   \int_{0}^\infty \Ab_n(x_1) \psi_n( x_1 - x) dx_1,
  \label{eqn:R-integral}
\end{equation}
where we assumed particles are distributed according to a uniform distribution, and the kernel $\psi_n$ is given by
\begin{equation}
  \psi_n(X) =  \int_{Y^2 > b^2 - X^2} \ee^{\ii \beta Y} (-1)^n \mathrm H^{(1)}_{n}(k R) \ee^{\ii n \Theta} dy.
  \label{eqn:integral_kernel}
\end{equation}
Later we show that, as expected, $\mathfrak R$ is independent of $x$.


The system governing $\Ab_m(x)$ is given by equation (4.7) of \cite{gower_multiple_2018}:
\begin{multline}
  \nfrac {} T_m \sum_{n=-\infty}^\infty
  \int_{0}^\infty \Ab_n (x_2) \psi_{n-m}(x_2 - x_1) \mathrm d x_2
  \\
   = \Ab_m(x_1) - \ee^{\ii \alpha x_1} T_m \ee^{\ii m ( \pi/2 - \theta_\inc )}, \quad \text{for} \;\; x_1  \geq 0,
  \label{eqn:Wiener-Hopf}
\end{multline}
for all integers $m$. Kristensson~\cite[equation (15)]{kristensson_coherent_2015} presents an equivalent integral equation for electromagnetism and particles in a slab.

Our main aim is to reach an exact solution for $\Ab_n(x)$ by employing the Wiener-Hopf technique to~\eqref{eqn:Wiener-Hopf}. We show how this also leads to simple solutions for the reflection coefficient by using~\eqref{eqn:AverageWave}. We acknowledge the authors of \cite{linton_multiple_2005}, as they noticed that~\eqref{eqn:Wiener-Hopf} is a Wiener-Hopf integral equation; but apparently did not follow the steps indicated in the following sections\footnote{However, they were unable to solve it because, it seems, of a mistake in the integrand of equation (37) of~\cite{linton_multiple_2005}; they used $\ee^{\ii \beta Y}$ where they should have used $\cos(\beta Y)$.}.

\section{Applying the Wiener-Hopf technique}
\label{sec:wiener-hopf}
Equation~\eqref{eqn:Wiener-Hopf} is convolution integral equation with a difference kernel. This means applying a Fourier transforms can lead to elegant and simple solutions. To facilitate, we must analytically extend~\eqref{eqn:Wiener-Hopf} for all $x_1 \in \mathbb R$ by defining
\begin{multline}
  \nfrac {} T_{m} \sum_{n=-\infty}^\infty
    \int_{0}^\infty  \Ab_n (x_2) \psi_{n-m}( x_2 - x_1) \mathrm d x_2
     \\
     =
  \begin{cases}
   \Ab_m(x_1) - \ee^{\ii \alpha x_1} T_m \ee^{\ii m ( \pi/2 - \theta_\inc )}, & x_1 \geq 0, \\
   \Db_m(x_1), & x_1 < 0,
  \end{cases}
  \label{eqn:Wiener-Hopf-extend}
\end{multline}
for integers $m$, where if the $\Ab_n(x)$ were known for $x >0$, then the $\Db_n(x)$ would be given from the left hand-side. Note that the kernel $\psi_n$ defined in \eqref{eqn:integral_kernel} is already analytic in the domain $\mathbb R$.

The field $\Db_0(x)$ is not just an abstract construct, it is closely related to the reflected wave: by directly comparing~\eqref{eqn:Wiener-Hopf-extend} with the reflection coefficient~\eqref{eqn:R-integral}, for $x < -b$, we find that
\begin{equation}
D_0(x) = T_0 \mathfrak R \ee^{-\ii \alpha x}.
\label{eqn:reflect-D0}
\end{equation}

To solve~\eqref{eqn:Wiener-Hopf-extend} we employ the Fourier transform and its inverse, which we define as
\begin{equation}
  \hat f(s) = \int_{-\infty}^\infty f(x) \ee^{\ii s x} dx \quad \text{with} \quad
  f(x) = \frac{1}{2\pi}\int_{-\infty}^\infty \hat f(s) \ee^{- \ii s x} dx,
  \label{eqn:FourierTransform}
\end{equation}
for any smooth function $f$. We then define
\begin{equation}
  \hat \Ab_n^+(s) = \int_{0}^\infty \Ab_n(x) \ee^{\ii  s x} dx, \quad
  \hat \Db_n^-(s) = \int_{-\infty}^0 \Db_n(x) \ee^{\ii s x} dx.
  \label{eqns:half_transforms}
\end{equation}

We can determine where $\hat \Ab_n^+$ and $\hat \Db_n^-$ are analytic by assuming\footnote{The solutions for $\Ab_n(x)$ and $\Db_n(x)$, in the next section, show that these assumptions do hold.} that
\begin{align}
  & |\Ab_n(x)| < \ee^{- x c } \;\; \text{for} \;\; x \to \infty,
  \label{ineq:V}
  \\
  & |\Db_n(x)| < \ee^{x c} \;\; \text{for} \;\; x \to -\infty,
  \label{ineq:D}
\end{align}
for some (possibly small) positive constant $c$. This leads to $\hat \Ab_n^+(s)$ being analytic for $\mathrm{Im} \, s >  - c$, while $\hat \Db_n^-(s)$ is analytic for $\mathrm{Im} \, s < c$. In other words, both~$\hat \Ab_n^+(s)$ and $\hat \Db_n^-(s)$ are analytic in the overlapping strip
\begin{equation}
|\mathrm{Im} \, s| <  c.
\label{eqn:transform_restriction}
\end{equation}

To apply the Wiener-Hopf technique we also need to specify the large $s$ behaviour for both $\hat \Ab_n^+(s)$ and $\hat \Db_n^+(s)$. To achieve this, we assume, on physical grounds, that $\Ab_n(x)$ is bounded when $x \to 0^+$, and $\Db_n(x)$ is bounded when $x \to 0^-$. Then, it can be shown \cite{noble_methods_1988,bleistein_asymptotic_1986} that
\begin{align}
  \hat \Ab_n^+(s) = \mathcal O(|s|^{-1}) \;\; \text{and} \:\; \hat \Db_n^-(s) = \mathcal O(|s|^{-1})    \quad \text{for} \;\; |s| \to \infty,
  \label{eqn:AbDb-decay}
\end{align}
in their respective half-planes of analyticity.


Applying a Fourier transform to both sides of equation~\eqref{eqn:Wiener-Hopf-extend}, the left-hand side becomes
\begin{equation}
   \int_{0}^\infty  \Ab_n (x_2) \int_{-\infty}^\infty \psi_{n-m}(x_1 - x_2) \ee^{\ii s x_1} \mathrm d x_1 \mathrm d x_2
   \\
  =
   \hat \Ab_n^+ (s) \hat \psi_{n-m}(s),
  \label{eqn:left-hand}
\end{equation}
in which $\hat \psi_n(s)$ is well defined (i.e.\ analytic) for $s$ in the strip:
\begin{equation}
  |\mathrm{Im} \, s | < (1 - |\sin \theta_\inc|) \mathrm{Im} \, k,
  \label{eqn:ImsImK}
\end{equation}
see~\cref{app:hat_psi} for details.
The right-hand side of \eqref{eqn:Wiener-Hopf-extend} becomes
\begin{multline}
  \int_{-\infty}^0 \Db_m(x_1) \ee^{\ii s x_1} dx_1 + \int_0^\infty \Ab_m(x_1) \ee^{\ii s x_1} \mathrm d x_1
  \\
   - \ee^{\ii m (\pi/2 - \theta_\inc)}T_m\int_0^{\infty} \ee^{\ii x_1 (s + \alpha)} \mathrm d x_1 =
   \hat \Db_m^- (s) + \hat \Ab_m^+(s)  -  T_m\frac{\ii \ee^{\ii m (\pi/2 - \theta_\inc)}}{(s +  \alpha)^+},
   \label{eqn:right-hand}
\end{multline}
where for the last step we assumed Im $(s + \alpha) > 0$, which is why we use the superscript $+$ on $(s +  \alpha)^+$. This assumption, together with~\eqref{eqn:transform_restriction} and \eqref{eqn:ImsImK}, is satisfied if
\begin{equation}
|\mathrm{Im}\, s| < \epsilon, \quad \text{where} \quad \epsilon = \min \{c, \, (1 - |\sin \theta_\inc|) \mathrm{Im} \, k, \, \mathrm{Im} \, \alpha \}.
\label{ineqn:s-strip}
\end{equation}

 If~\eqref{ineqn:s-strip} is satisfied then we can combine~\eqref{eqn:left-hand},\eqref{eqn:right-hand} and \eqref{eqn:KernalFourierResult}, to obtain the Fourier transform of \eqref{eqn:Wiener-Hopf-extend} in matrix form:
\begin{equation}
  \frac{\vec \Psi(s)  \hat {\vec \Ab}^+ (s)}{s^2 - \alpha^2}   =  - \hat{ \vec \Db}^- (s)   + \frac{\vec B}{s + \alpha},
  \label{eqn:WH-Fourier-Matrix}
\end{equation}
 where $\hat {\vec \Ab}^+(s)$ and $\hat{ \vec \Db}^- (s)$ are vectors with components $\hat \Ab_n^+ (s)$ and $\hat \Db_n^- (s)$, respectively and
 \begin{align}
   & B_m = \ii T_m \ee^{\ii m (\pi/2 - \theta_\inc)},
   \label{eqn:b_comp}
   \\
   & \Psi_{mn} (s) =  G_{mn} ( S)(-\ii)^{n-m} \ee^{ \ii (n - m) \theta_S},
   \label{eqn:PsiToG}
   \\
   & G_{mn}(S) =
   (s^2 - \alpha^2) \delta_{mn}  +  2 \pi \nfrac {} T_m \mathrm N_{n-m}(b S),
   \label{eqn:F_comp}
   \\
   & \mathrm N_{m}(b S) =
   b k \mathrm J_{m}(b S) \mathrm H^{(1)\prime}_{m}(b k) - b S \mathrm J_{m}'(b S) \mathrm H^{(1)}_{m}(b k).
   \label{eqn:N_comp}
\end{align}
where, for reference,
\begin{equation}
\Psi_{mn}(s) = (s^2-\alpha^2) \left[ \delta_{mn} - \nfrac {} T_m \hat \psi_{n-m}(s) \right ],
\label{eqn:psiToPsi}
\end{equation}
 and $\hat \psi_{n-m}(s)$ is given by \eqref{eqn:KernalFourierResult}.
In the above $\theta_S$ and $S$ are chosen to satisfy
\begin{equation}
  s = S \cos \theta_S \quad \text{with} \quad S \sin \theta_S = k \sin \theta_\inc.
  \label{eqn:SthetaS}
\end{equation}
Later we identify $S$ and $\theta_S$ as the effective wavenumber and transmission angle. The above does not determine the sign of $S$ for any given complex $s$. To fully determine $S$ and $\theta_S$, we take $\sgn(\mathrm{Re}\,s) = \sgn(\mathrm{Re}\,S)$
 which together with~\eqref{eqn:SthetaS} leads to
\begin{align}
  & \theta_S = \arctan \left(\frac{k \sin \theta_\inc}{s }\right), \quad  S =
        \sqrt{s^2 + (k \sin \theta_\inc)^2},
  \label{eqn:sToS}
\end{align}
where both $S$ and $\theta_S$, when considered as functions\footnote{With our choice of branch-cut, the simplest way to compute $S$, with most software packages, is to use $S = \sgn(\mathrm{Re}\, s) \sqrt{s^2 + (k \sin \theta_\inc)^2}$ with the default cut location for the square root function, i.e.\ along the real negative line.} of $s$, contain branch-points at $s = \pm \ii k \sin \theta_\inc$ with finite branch-cut running between $-\ii k \sin \theta_\inc$ and $\ii k \sin \theta_\inc$.
     However, $\vec \Psi(s)$ is an entire matrix function having only zeros in $s$ and no branch-points; see the end of \cref{app:hat_psi} for details.



Determining the roots of $\det \vec \Psi ({s}) =0$ will be a key step in solving~\eqref{eqn:WH-Fourier-Matrix}, and so the following identities will be useful
\begin{align}
  & \Psi_{mn}(-{s}) T_n = \Psi_{mn}({s}) T_n (-1)^{m-n} \ee^{2 \ii (m - n) \theta_s} = \Psi_{nm}({s}) T_m,
  \label{eqn:PsiSign}
  \\
  & \det \vec \Psi(-{s}) = \det \vec \Psi({s}) \quad \text{and}
  \quad \det \vec \Psi(s) = \det \mathbf G (S).
  \label{eqn:Fproperties}
\end{align}
where \eqref{eqn:PsiSign} results from \eqref{eqn:psiToPsi} and~\eqref{eqn:psi-properties}. Equation~\eqref{eqn:Fproperties} then follows from using~\eqref{eqn:PsiSign}${}_1$, \eqref{eqn:PsiToG}, and \cref{app:formulas}.


\subsection{Multiple waves solution}
To solve~\eqref{eqn:WH-Fourier-Matrix}, we use a matrix product factorisation~\cite{gokhberg_systems_1960} of the form:
\begin{equation}
  \vec \Psi({s}) = \vec \Psi^-({s}) \vec \Psi^+({s}),
  \label{eqn:F-Split}
\end{equation}
where $\vec \Psi^-({s})$, and its inverse, are analytic in Im $s < \epsilon$, and $\vec \Psi^+({s})$, and its inverse, are analytic for Im $s > -\epsilon$. See~\eqref{ineqn:s-strip} for the definition of $\epsilon$.


For our purposes, it is enough to know that such a factorisation exists~\cite{gokhberg_systems_1960}, as this will lead to a proof that $\vec \Ab(x)$ is a sum of attenuating plane waves.

Multiplying both sides of \eqref{eqn:WH-Fourier-Matrix} by $[\vec \Psi^-({s})]^{-1}$  and by $({s} - {\alpha})_-$ leads to
\begin{equation}
  \frac{\vec \Psi^+({s}) \hat {\vec \Ab}^+ ({s})}{({s} + {\alpha})_+}   =  -({s} - {\alpha})_-[\vec \Psi^-({s})]^{-1} \hat{\vec \Db}^- ({s})
  + [\vec \Psi^-({s})]^{-1} \vec B \frac{({s} - {\alpha})_-}{({s} + {\alpha})_+},
  \label{eqn:pre-factorise}
\end{equation}
where $(s + \alpha)_{+}$ is analytic for $\mathrm{Im}\,s > -\mathrm{Im}\,\alpha$, while $(s - \alpha)_{-}$ is analytic for $\mathrm{Im}\,s < \mathrm{Im}\,\alpha$. We need to rewrite the last term above as a sum of a function which is analytic in the upper half-plane (Im $s> -\epsilon$) and another analytic in the lower half-plane. This is achieved below:
\begin{multline}
   [\vec \Psi^-(s)]^{-1} \vec B \frac{({s} - {\alpha})_-}{({s} + {\alpha})_+} = - \underbrace{
     \frac{2 \alpha}{({s} + {\alpha})_+} [\vec \Psi^-(-\alpha)]^{-1} \vec B
     }_{ \vec g^+(s)}
 \\
   + \underbrace{[\vec \Psi^-(s)]^{-1} \vec B \frac{({s} - {\alpha})_-}{({s} + {\alpha})_+} + [\vec \Psi^-(-\alpha)]^{-1} \vec B \frac{2 \alpha }{({s} + {\alpha})_+}}_{\vec g^-(s)},
   \label{eqn:gFactors}
\end{multline}
where we define
\begin{equation*}
  \lim_{s \to - \alpha} \vec g^-(s) = \left [ \mathbf I + 2 \alpha [\vec \Psi^-(-\alpha)]^{-1} \frac{\mathrm d \vec \Psi^-}{\mathrm d s}(-\alpha) \right ][\vec \Psi^-(- \alpha)]^{-1} \vec B,
\end{equation*}
so that $\vec g^-(s)$ does not have a pole at $s = - \alpha$ and is therefore analytic for $\mathrm{Im} \, s < \epsilon$.

 Substituting~\eqref{eqn:gFactors} into~\eqref{eqn:pre-factorise} leads to
\begin{equation}
  \frac{\vec \Psi^+(s) \hat {\vec \Ab}^+ (s)}{(s + \alpha)_+} + \vec g^+(s)  =  -(s - \alpha)_-[\vec \Psi^-(s)]^{-1} \hat{ \vec \Db}^- (s)
  + \vec g^-(s).
  \label{eqn:entire-E}
\end{equation}
Because both sides are analytic in the strip $|\mathrm{Im}\, s| < {\epsilon}$, we can equate each side to $\vec E(s)$, some analytic function in the strip. Further, as the left-hand side (right-hand side) of \eqref{eqn:entire-E} is analytic for $\mathrm{Im}\, s > \epsilon$ ( $\mathrm{Im}\, s < -\epsilon$), we can analytically continue $\vec E(s)$ for all $s$, i.e.\ $\vec E(s)$ is entire.

To determine $\vec E(s)$ we need to estimate its behaviour as $|s| \to \infty$. From~\eqref{eqn:AbDb-decay} we have that $\Ab^+(s) = \mathcal (|s|^{-1})$ as $|s| \to \infty$ in the upper half-plane, and from~(\ref{eqn:PsiToG} - \ref{eqn:N_comp}):
\begin{equation}
  \vec \Psi(s) = (s^2 - \alpha^2) \vec I + \mathcal O(|s|) \quad \text{as} \quad |s| \to \infty,
\end{equation}
for $s$ in the strip~\eqref{ineqn:s-strip}. From this we know that the factors $\vec \Psi^+(s)$ and $\vec \Psi^-(s)$ must be $\mathcal O(|s|)$ as $|s| \to \infty$, in their respective half-planes of analyticity~\cite{abrahams_solution_1997}. So, the left hand-side of~\eqref{eqn:entire-E} behaves as $\mathcal O(|s|^{-1})$ as $|s| \to \infty$ in $\mathrm{Im}\,s > - \epsilon$. We can therefore use Liouville's theorem to conclude that $\vec E(s) \equiv 0$, which means the Wiener-Hopf equation~\eqref{eqn:entire-E} is formally equivalent to
\begin{align}
    \hat {\vec \Ab}^+ (s)  = & - 2 \alpha [\vec \Psi^+(s)]^{-1} [\vec \Psi^-(-\alpha)]^{-1} \vec B,
    \label{eqn:V+}
  \\
  \hat{ \vec \Db}^- (s) = & \frac{\vec \Psi^-(s) \vec g^-(s)}{(s - \alpha)_-}.
 \label{eqn:D-}
\end{align}
Let $\mathbf C^+(s)$ be the cofactor matrix of $\vec \Psi^+(s)$, so that
\[
[\vec \Psi^+(s)]^{-1} = \frac{[\mathbf C^+(s)]^\mathrm{T}}{\det(\vec \Psi^+(s))}.
\]
From the property~\eqref{eqn:Fproperties}${}_1$ we can write $\det \vec \Psi(s) = f(s^2)$ for some function $f$. Then, for every root $s = s_p$ of $\det \vec \Psi(s)$, with Im $s_p >0$, we have that $-s_p$ is also a root, and vice-versa. From here onwards we assume:
\begin{equation}
  \det \vec \Psi(s_p) = \det \vec \Psi(-s_p) = 0 \quad \text{with} \quad \mathrm{Im}\, s_p >0 \quad \text{and} \quad p=1,2,\cdots,\infty.
  \label{eqn:dispersion}
\end{equation}
For any truncated matrix $\vec \Psi(s)$, i.e.\ evaluating $m,n= -M,\ldots,M$ in~\eqref{eqn:PsiToG}, the roots $s_p$ are discrete. In \cref{sec:multiple-wavenumbers} we demonstrate asymptotically that they are indeed discrete for the limits of low and high wavenumber $k$.
For the numerical results presented in this paper, we numerically solve the above dispersion relation for the truncating the matrix $\vec \Psi(s)$, and then increase $M$ until the roots converge (typically no more than $M =4$ was required).

Given $\det \vec \Psi(s) = \det \vec \Psi^-(s) \det \vec \Psi^+(s)$, every root of $\det \vec \Psi(s)$ must either be a root of $\det \vec \Psi^-(s)$ or a root of $\det \vec \Psi^+(s)$. For $[\vec \Psi^+(s)]^{-1}$ to be analytic in the upper half-plane, $\det \vec \Psi^+(s)$ must only have roots $s = -s_p$. As a consequence, $\det \vec \Psi^-(s)$ only has roots $s = s_p$.

To use the residue theorem below, we need to calculate $\det \vec \Psi^+(s)$ for $s$ close to the root $-s_p$, in the form
\begin{multline}
\det \vec \Psi^+(s) = \det \vec \Psi^+(-s_p) +  (s + s_p) \frac{\mathrm d \det \vec \Psi^+}{\mathrm d s}(-s_p) + \mathcal O((s + s_p)^2)
\\
= \frac{s + s_p}{\det \vec \Psi^-(-s_p)} \frac{\mathrm d \det \vec \Psi}{\mathrm d s}(-s_p)  + \mathcal O((s + s_p)^2)
\end{multline}
where we use $\frac{d \det \vec \Psi}{ds}(-s_p)$ instead of $\frac{d \det \vec \Psi^+}{ds}(-s_p) \det \vec \Psi^-(-s_p)$, because it is more difficult to numerically evaluate $\frac{d \det \vec \Psi^+}{ds}(-s_p)$.

\begin{figure}[ht!]
  \centering
  \includegraphics[width=0.7\linewidth]{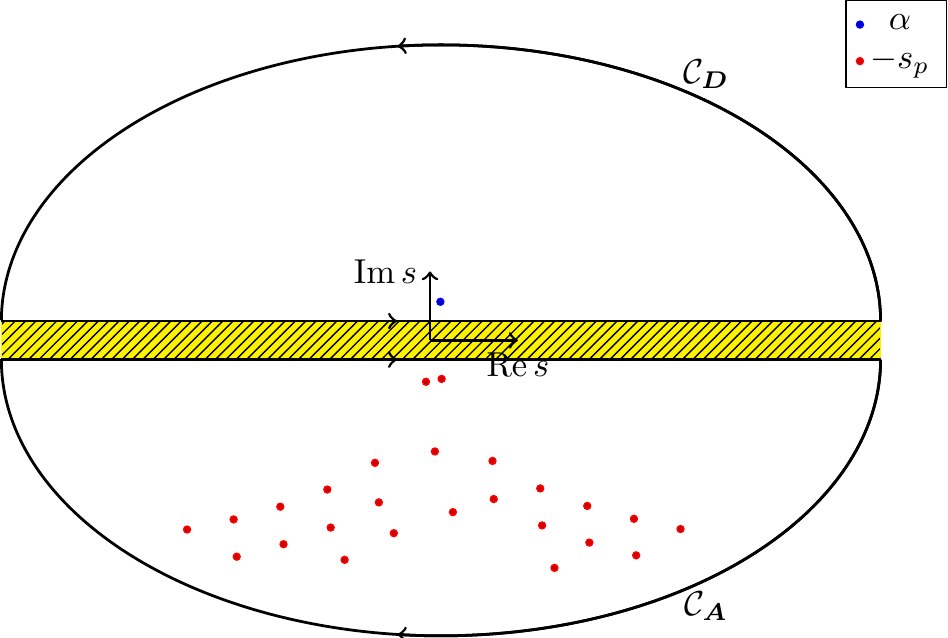}
  \caption{An illustration of the contour integral over $\mathcal C_{\vec D}$, used to calculate~\eqref{eqn:D-matrix} for $x<0$, and the contour integral over $\mathcal C_{\vec \Ab}$, used to calculate \eqref{eqn:V-matrix} for $x>0$. The $-s_p$ (the red points) are roots of~\eqref{eqn:dispersion}, and also the poles of~\eqref{eqn:V+}. The single blue point $\alpha$ is the only pole of~\eqref{eqn:D-}.}
  \label{fig:integral_paths}
\end{figure}

Using the above, and that $\mathbf C^+(S)$ is analytic for Im $s > - {\epsilon}$, we can apply an inverse Fourier transform~\eqref{eqn:FourierTransform}${}_2$ to both sides of \eqref{eqn:V+} and using residue calculus we find
\begin{align}
   & \vec \Ab(x) = - \frac{\alpha}{\pi}\int_{-\infty}^\infty
  \frac{[\mathbf C^+(s)]^\mathrm{T} [\vec \Psi^-(-\alpha)]^{-1}\vec B}{\det \vec \Psi^+(s)}
     \ee^{-\ii {s} x} \mathrm d s
     = \begin{cases}
     \sum_{p=1}^\infty \vec \Ab^p
       \ee^{\ii s_p x}, &  x >0,
       \\
       0, & x < 0,
     \end{cases}
       \label{eqn:V-matrix}
       \\
    & \text{with} \qquad  \vec \Ab^p  =  2 \alpha \ii \frac{\det \vec \Psi^-(-s_p)}{\frac{\mathrm d \det \vec \Psi}{\mathrm d s}(-s_p) } [\mathbf C^+(-s_p)]^\mathrm{T}
          [\vec \Psi^-(-\alpha)]^{-1}\vec B.
\end{align}
For $x>0$, the integral over $s \in [-\infty,\infty]$ in~\eqref{eqn:V-matrix} is, by Jordan's lemma, the same as a clockwise integral over the closed contour $\mathcal C_{\vec \Ab}$ which surrounds the poles $-s_1,\, -s_2, \ldots $, i.e.\ roots of~\eqref{eqn:dispersion}, as shown by \Cref{fig:integral_paths}.
 Note that the cofactor matrix $\mathbf C^+(s)$ contains no poles and so does not contribute additional residual terms.
 The yellow striped region in \Cref{fig:integral_paths} is the domain where $\vec \Psi$ is analytic. On the other hand, for $x<0$, the integral~\eqref{eqn:V-matrix} is the same as an integral over the counter-clockwise closed contour within the region Im $s > 0$ (not shown in \Cref{fig:integral_paths}). The integrand has no poles in this domain and hence evaluates to zero.

Likewise, by applying an inverse Fourier transform to ~\eqref{eqn:D-}, we obtain:
\begin{equation}
  \vec \Db(x) = \frac{1}{2 \pi} \int_{-\infty}^\infty \frac{\vec \Psi^-(s) \vec g^-(s)}{(s - \alpha)_-} \ee^{-\ii s x} \mathrm d s
  = \begin{cases}
    \ii \vec \Psi^-(\alpha) [\vec \Psi^-(-\alpha)]^{-1} \vec B  \ee^{-\ii \alpha  x}, & x<0,
    \\
    0, & x>0,
  \end{cases}
  \label{eqn:D-matrix}
\end{equation}
For $x < 0$ the above integral is the same as a counter-clockwise closed integral over $\mathcal C_{\vec \Db}$ which surrounds the pole $s = \alpha$ (recalling that Im $\alpha>0$), as shown in \Cref{fig:integral_paths}. The result is just the residue at this pole. That is, the function $\vec \Psi^-(s) \vec g^-(s)$ contains no other singularities within $\mathrm{Im}\,s > 0$. On the other hand, for $x>0$ the integral is the same as a closed clockwise integral around the region Im $s <0$ which evaluates to zero, as there are no singularities in this region (not shown in \Cref{fig:integral_paths}).

Clearly~\eqref{eqn:V-matrix} shows that $\vec \Ab(x)$ is a sum of plane waves with different effective wavenumbers $s_p$, each satisfying~\eqref{eqn:dispersion}. In \cref{sec:multiple-wavenumbers} we discuss these roots in more detail, and in
  \cref{sec:Numerical}, we see that usually only a few effective wavenumbers are required to obtain accurate results.

\subsection{Reflection coefficient}
By substituting~\eqref{eqn:D-matrix} in~\eqref{eqn:reflect-D0} leads to
\begin{equation}
  \mathfrak R =  \ii T_0^{-1} \sum_{n,m=-\infty}^\infty \Psi^-_{0n}(\alpha) [\vec \Psi^-(-\alpha)]_{nm}^{-1} B_m .
  \label{eqn:R}
\end{equation}
Alternatively, the reflection coefficient can be calculated from~\eqref{eqn:R-integral} by employing the form of $\vec \Ab(x)$ from \eqref{eqn:V-matrix}, which is the more common approach. To simplify, we use
\begin{align}
  & \psi_n(X) = (-1)^n \int_{-\infty}^\infty \ee^{\ii k Y \sin \theta_\inc}\mathrm H^{(1)}_n(k R) \ee^{\ii n \Theta} dY = \frac{2}{\alpha} \ii^n \ee^{-\ii n \theta_\inc}\ee^{\ii \alpha X} \quad \text{for} \;\; X > 0,
\end{align}
which then implies that $\psi_n(x_1 - x) = \frac{2}{\alpha} \ii^n \ee^{-\ii n \theta_\inc}\ee^{\ii \alpha (x_1 - x)}$ for $x_1 \geq x$. The above is shown in \cite[equation (37)]{martin_multiple_2008} and \cite[equation (65)]{linton_multiple_2005}. This result together with \eqref{eqn:V-matrix} substituted into \eqref{eqn:R-integral} leads to the form
\begin{equation}
  \mathfrak R =  \frac{2 \nfrac {}}{{\alpha}} \sum_{n=-\infty}^\infty \ii^n \ee^{-\ii n \theta_\inc}  \int_{0}^\infty \Ab_n(x_1) \ee^{\ii {\alpha} x_1} dx_1
  = \frac{2 \ii \nfrac {}}{\alpha} \sum_{n=-\infty}^\infty \sum_{p=1}^\infty  \ii^n \ee^{-\ii n \theta_\inc}
  \frac{\Ab^p_n}{s_p + \alpha},
  \label{eqn:R_P}
\end{equation}
where we used that Im $s_p >0$. The above agrees with \cite[equation (39)]{martin_multiple_2011} and\footnote{When taking a zero thickness boundary layer, i.e. $J=0$, and appropriate substitutions.}~\cite[equation (6.9)]{gower_multiple_2018}.

\section{Monopole scatterers}
\label{sec:monopole}
For particles that scatter only in their monopole mode, i.e. the scattered waves are angularly symmetric about each particle, we can easily calculate the factorisation~\eqref{eqn:F-Split}. This type of scattered wave tends to dominate in the long wavelength limit for scatterers with Dirichlet boundary conditions. In acoustics, these correspond to particles with low density or low sound speed.


Once we know the factorisation~\eqref{eqn:F-Split}, we can then calculate the average scattering coefficient~\eqref{eqn:V-matrix} and average reflection coefficient~\eqref{eqn:R}. We will compare both of these against predictions from other methods in~\cref{sec:Numerical}.

\subsection{Wiener-Hopf factorisation}
For scalar problems, there are well known techniques to factorise $\Psi_{00}(s) = \Psi_{00}^-(s) \Psi_{00}^+(s)$, such as Cauchy's integral formulation, for details see~\cite[Section 5. Wiener-Hopf Technique]{crighton_modern_1992} and \cite{noble_methods_1988}.

For monopole scatterers we use $S^2 - k^2 = s^2 - \alpha^2$ and rewrite
\[
\Psi_{00}(s)= (s^2 - \alpha^2) q(s), \quad \text{with}\;\;  q(s) =
1  +
2 \pi \nfrac {} \frac{T_0 \mathrm N_{0}(b S)}{S^2 - k^2},
\]
with $\mathrm N_0(b S)$ given by~\eqref{eqn:N_comp}. Then, because $q(s) \to 1$ as $|s| \to \infty$, we can factorise $q(s) = q^-(s) q^+(s)$ using
\begin{align}
  q^+(s) =& \exp\left( \frac{1}{2 \pi \ii}\landdownint_{-\infty}^{\infty} \frac{\log q(z)}{z-s} dz\right),
  \label{eqn:q-plus}
  \\
  q^-(s) =& \exp\left( -\frac{1}{2 \pi \ii}\landupint_{-\infty}^{\infty} \frac{\log q(z)}{z-s} dz\right),
  \label{eqn:q-minus}
\end{align}
where the integral path for $q^+(s)$ ($q^-(s)$) has to be in the strip where $q(s)$ is analytic, with the path for $q^+(s)$ ($q^-(s)$) passing below (above) $z$.
 We then have\footnote{Note that the factors $q^+(s)$ and $q^-(s)$ are singularity and pole free in their respective regions of analyticity, and so their inverses $[q^+(s)]^{-1}$ and $[q^-(s)]^{-1}$ have the same property.} that
\begin{equation}
  \Psi^+_{00}(s) = (s + \alpha)_+ q^+(s), \;\; \Psi^-_{00}(s) = (s - \alpha)_- q^-(s), \;\; \Psi^-_{00}(-s) = - \Psi^+_{00}(s),
  \label{eqn:split_sign_change}
\end{equation}
where~\eqref{eqn:split_sign_change}${}_3$ holds if $-s$ is below the integration path of~\eqref{eqn:q-minus} and $s$ is above the integration path of~\eqref{eqn:q-plus}.
 From~\eqref{eqn:V-matrix} we see that we need only evaluate $\Psi^+_{00}(s)$, and therefore $q^+(s)$, for $s=s_1,s_2,\ldots, s_p$ where as $p$ increases, the $s_p$ become more distant from the real line. Then for large $z$, by inspection of~\eqref{eqn:N_comp}, we have that
 \[
\left | \frac{\log q(z)}{z-s} \right | \sim \frac{1}{z^{3/2}} \frac{1}{|z-s|}, 
 \]
and therefore we can accurately approximate the integral~\eqref{eqn:q-plus} by truncating the integration domain for large $z$.

\subsection{Explicit solution for monopole scatterers.}

For monopole scatterers $\Ab_n(x) = \Db_n(x) = 0$ for $|n|>0$. Using this in~(\ref{eqn:b_comp} - \ref{eqn:N_comp}) leads to all vectors and matrices having only one component, given by setting $n=m=0$. In this case $\vec \Ab$~\eqref{eqn:V-matrix} reduces to
\begin{equation}
  \Ab_{0}(x) = \sum_{p=1}^\infty \Ab_0^p \ee^{\ii s_p x} \quad \text{with} \quad
  \Ab_0^p = \frac{2 \alpha T_0}{\Psi^+_{00}(\alpha)}
     \frac{\Psi^+_{00}(s_p)}{ \frac{\mathrm d \Psi_{00}}{\mathrm d s}(s_p)} =
     \frac{T_0}{s_p - \alpha}
        \frac{q^+(s_p)}{q^+(\alpha) q'(s_p)},
     \label{eqn:V-0}
\end{equation}
for $x>0$, where we used \eqref{eqn:split_sign_change}, $\mathbf C^+(s) = 1$, $\vec B = \ii T_0$,  and $\frac{d\Psi_{00}}{d s}(-s) = - \frac{d\Psi_{00}}{d s}(s)$ for every $s$.
Likewise for~\eqref{eqn:R} we arrive at
\begin{equation}
  \mathfrak R  = \frac{\Psi_{00}(\alpha)}{(\Psi^+_{00}(\alpha))^2} = \frac{\pi \nfrac {} T_0 \mathrm N_0(b \alpha)}{2 (\alpha q^+(s))^2}.
  \label{eqn:D-0}
\end{equation}

Alternatively, using~\eqref{eqn:R_P}, we can calculate the contribution of $P$ effective waves to the reflection coefficient
\begin{equation}
  \mathfrak R^P = \frac{2 \ii \nfrac {}}{\alpha} \sum_{p=1}^P \frac{\Ab^p_0}{s_p + \alpha}
  = \frac{2 \ii \nfrac {} T_0}{\alpha q^+(\alpha)} \sum_{p=1}^P \frac{1}{s_p^2 - \alpha^2}\frac{q^+(s_p)}{q'(s_p)}
  \quad \text{with} \quad \mathfrak R = \lim_{P \to \infty} \mathfrak R^P,
  \label{eqn:R_P_0}
\end{equation}
where the error $|\mathfrak R^P - \mathfrak R|$ then indicates how many effective waves are needed to accurately describe the field near the boundary $x=0$.

\begin{figure}[ht]
\centering
\input{"images/wavenumbers-example.tex"}
\caption{Examples fo effective wavenumbers $S_p$ which satisfy the dispersion equation~\eqref{eqn:detF} with the properties~\eqref{eqn:parameters}. The blue points represent waves travelling forwards (i.e.\ deeper into the material), while the red represent waves travelling backwards. All these waves are excited in a reflection experiment. Two wavenumbers in particular stand out as having the lowest attenuation $S_1$ and $S_2$, both inside the grey dashed circle. The graph on the right is a magnification of the region close to these two wavenumbers. Out of these two, most efforts in the literature have focused on calculating $S_2$, as it often has the lowest imaginary part; however for this case, because $S_1$ has a smaller attenuation it will have a significant contribution to both transmission and reflection.}
\label{fig:wavenumbers-example}
\end{figure}
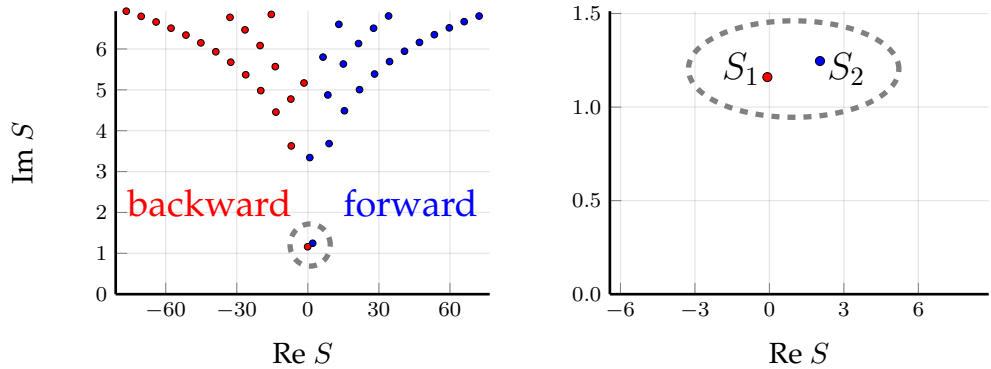

\section{Multiple effective wavenumbers}
\label{sec:multiple-wavenumbers}

Equation~\eqref{eqn:V-matrix} clearly shows that $\vec \Ab(x)$ is a sum of attenuating plane waves, each with a different effective wavenumber $s_p$. These $s_p$ satisfy the dispersion equation~\eqref{eqn:dispersion}:
\begin{equation}
  \det \vec \Psi(s_p) = \det \mathbf G(S_p) = 0,
  \label{eqn:detF}
\end{equation}
with $\vec \Psi$ given by~\eqref{eqn:F_comp} and the first identity follows from~\eqref{eqn:Fproperties}.

An important conclusion from $\det \mathbf G(S_p) = 0$ is that the wavenumbers $S_p$ are independent of the angle of incidence $\theta_\inc$. We focus on showing the results for $S_p$, rather than $s_p$, because then we do not need to specify $\theta_\inc$.

As a specific example, let us consider circular particles with Dirichlet boundary conditions (i.e.\ particles with zero density or soundspeed), and the parameters
\begin{equation}
  T_n = - \frac{\mathrm J_n(k a_o)}{\mathrm H^{(1)}_n(k a_o)}, \quad  k b = 1.001, \quad k a_o = 0.5, \quad \phi = 30\%,
  \label{eqn:parameters}
\end{equation}
where $a_o$ is the radius of the particle.

With the above parameters, we found that truncating the matrix $\vec \Psi(s)$, with $|n| \leq 3$ and $|m| \leq 3$ in (\ref{eqn:PsiToG}-\ref{eqn:N_comp}), led to accurate results when calculating the effective wavenumbers $S_p$, i.e. the roots of~\eqref{eqn:dispersion}. Numerically calculating the wavenumbers $S_p$ then leads to~\Cref{fig:wavenumbers-example}.

The effective wavenumbers with the lowest attenuation (smallest imaginary part) contribute the most to the transmitted wave. In~\Cref{fig:wavenumbers-example} we see two wavenumbers have lower attenuation then the rest, both within the dashed grey circle. The blue point represents the wavenumber that most of the literature focuses on calculating: it has a positive real part and therefore propagates forwards along the $x-$axis (into the material) as is expected for a transmitted wave. However, the other wavenumber, with negative real part, is equally as important because it actually has lower attenuation. \Cref{fig:multiple-diagram} illustrates several effective wavenumbers, some travelling forward into the material, while others have negative phase direction (travel backwards).

In~\Cref{fig:wavenumbers-example} we see what appears to be an infinite sequence of effective wavenumbers $S_p$, where $|S_p| \to \infty$ as $p \to \infty$. To confirm their existence, and to find their locations as $|p| \to \infty$, we develop asymptotic formulas~in \cref{app:asymptotic-wavenumbers}. The results of the asymptotics are summarised below.

\begin{figure}[ht!]
  \centering
  \includegraphics[width=0.7\linewidth]{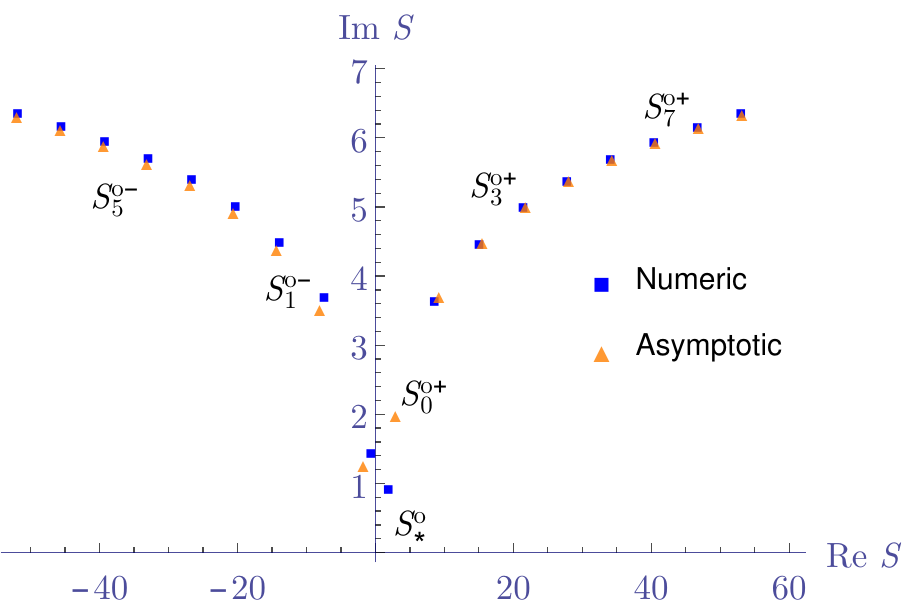}
  \caption{Comparison of the asymptotic formula~\eqref{eqn:asymS_p}, which predicts an infinite number of effective wavenumbers, with numerical solutions for the effective wavenumbers~\eqref{eqn:detF}. The parameters used are given by \eqref{eqn:parameters}, with their definitions explained in (\ref{eqn:incident}--\ref{eqn:non-dimensional}). Here we chose $b = 1.0$, so the non-dimensional wavenumbers $b S$ are the same as shown. The asymptotic formula is surprisingly accurate except for the two lowest attenuating wavenumbers. The wavenumber $S_*^o$ can be calculated by using low volume fraction expansions~\cite{linton_multiple_2005}.}
  \label{fig:AnalyticWavenumbers}
\end{figure}

For \emph{monopole scatterers}, where $n=m =0$ in~\eqref{eqn:PsiToG}, equations~\eqref{eqns:monopole-asym} give the effective wavenumbers $S_p^o$ at leading order:
\begin{align}
  & b S_p^{o\pm} =\sigma_p^\pm + \ii \log \left(\frac{|\sigma_p^\pm|^{3/2}}{r_c} \right),
\quad
 \begin{cases}
  \sigma_p^+ =   \theta_c + 2\pi p \;\; &\text{for} \;\;  p > -\left \lceil \frac{\theta_c}{2 \pi} \right \rceil,
  \\
     \sigma_p^- = \theta_c - \frac{3 \pi}{2} - 2\pi p  \quad &\text{for} \;\;   p > \left \lceil \frac{\theta_c}{2\pi} - \frac{3}{4}  \right \rceil,
\end{cases}
  \label{eqn:asymS_p}
  \\
  &
  r_c \ee^{\ii \theta_c} = \sqrt{2\pi} \nfrac {} b^2 T_0 \mathrm H^{(1)}_0(k b)\ee^{-\frac{\ii \pi }{4}}, \quad r_c > 0, \quad -\pi \leq \theta_c \leq \pi,
\end{align}
and for any integer $p$. We use the superscript ``o'' to distinguish these wavenumbers for monopole scatterers from others. Even though~\eqref{eqn:asymS_p} was deduced for large integer $p$, it gives remarkably agreement with numerically calculated wavenumbers, except for the two lowest attenuating wavenumbers, as shown in~\Cref{fig:AnalyticWavenumbers}. In the figure we denoted $S_*^{o\pm}$ as the effective wavenumber that can be calculated by low volume fraction expansions~\cite{linton_multiple_2005,norris_multiple_2011}.

For \emph{multipole scatterers}, where both $n$ and $m$ could potentially range from $-\infty$ to $\infty$ in~\eqref{eqn:PsiToG}, we can also calculate an infinite sequence of effective wavenumbers. To show this explicitly, we consider the limit of large $b k$, with $|k| \sim |S|$. In the opposite limit $b k \ll 1$, the Rayleigh limit, only one effective wavenumber is required~\cite{parnell_multiple_2010,parnell_effective_2010}.

At leading order, the asymptotic solution of~\eqref{eqn:pre-SMasym} leads to the effective wavenumbers:
\begin{align}
& b S_p^{k\pm} =\sigma_p^\pm + \ii \log \left(\frac{|\sigma_p^\pm - a|\sqrt{a |\sigma_p^\pm|}}{r_c} \right),
\label{eqn:SMasym}
\\
& \begin{cases}
  \sigma_p^+ =   \theta_c + a + 2\pi p \;\; &\text{for} \;\;  p > \left \lceil -\frac{\theta_c + a}{2 \pi} \right \rceil,
  \\
     \sigma_p^- = \theta_c + a - \frac{3 \pi}{2} - 2\pi p  \quad &\text{for} \;\;   p > \left \lceil \frac{\theta_c + a}{2\pi} - \frac{3}{4}  \right \rceil,
\end{cases}
\\
& r_c \ee^{\ii \theta_c} = - 2\ii \nfrac {} b^2  \sum_{n=-\infty}^\infty T_n,
\;\; r_c > 0 \;\; \text{and} \;\; -\pi \leq \theta_c \leq \pi,
\end{align}
for integer $p$. This confirms that there are an infinite number of effective wavenumbers for large scatterers, i.e. $b k\gg 1$. The distribution of these wavenumbers is similar to the monopole wavenumbers shown in~\Cref{fig:AnalyticWavenumbers}.

These asymptotic formulas~\eqref{eqn:asymS_p} and~\eqref{eqn:SMasym} demonstrate the existence of multiple effective waves in the limit of small (monopole and Dirichlet) scatterers~\eqref{eqn:asymS_p} and large scatterers~\eqref{eqn:SMasym}. However, neither of these formulas, nor the low volume fraction expansions of the wavenumber~\cite{linton_multiple_2005}, are able to accurately estimate the low attenuating backward travelling effective wavenumber such as $S_1$ shown in~\Cref{fig:wavenumbers-example} (in this case not related to the $S_1^{o\pm}$ and $S_1^{k\pm}$ given above).
 There is currently no way to analytically estimate these types of wavenumbers, even though they are necessary to accurately calculate transmission due to their small attenuation. The only approach it seems is to numerically solve~\eqref{eqn:dispersion}.

%
%

\section{Numerical results}
\label{sec:Numerical}
Here we present numerical results for monopole scatterers, as these have explicit expressions for reflection~\eqref{eqn:D-0} and the transmitted wave~\eqref{eqn:V-0} (or more accurately the average scattering coefficients). We compare our analytic solution with a classical method that assumes only one effective wavenumber~\cite{linton_multiple_2005,martin_multiple_2011}, and the Matching Method~\cite{gower_multiple_2018}, recently proposed by the authors. It should be noted that all of these approaches aim to solve the same equation~\eqref{eqn:Wiener-Hopf}.

Note that for monopole scatterers, using only one effective wavenumber $s_1$ can, in some cases, lead to accurate results. However, for multipole scatterers (a more common scenario practically) this is rarely the case because, as shown by \Cref{fig:wavenumbers-example}, there can be at least two effective wavenumbers with low attenuation, and therefore both are needed to obtain accurate results.

\begin{figure}[htb]
  \centering
  \includegraphics{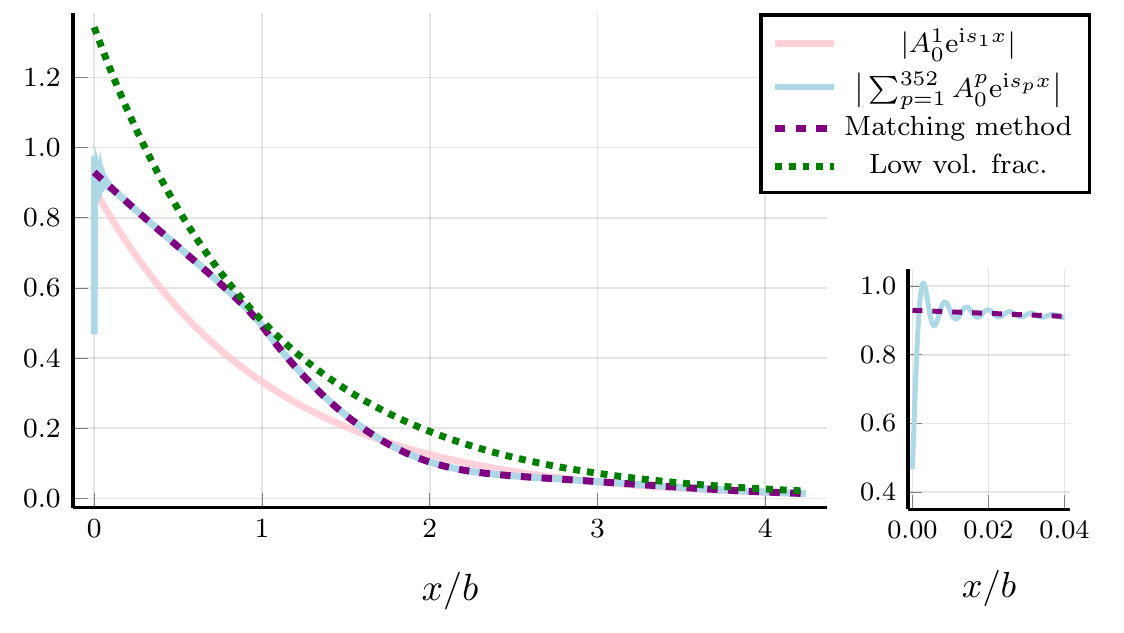}
  \caption{Compares the absolute value of the average field $\Ab_0(x)$ calculated by different methods. The field $\Ab_0(x)$ is closely related to the average transmitted wave~\cite{martin_multiple_2011}. The non-dimensional wavenumber $k b = 1.001$, the other parameters are given by~\eqref{eqn:parameters2}, with their definitions explained by (\ref{eqn:incident}--\ref{eqn:non-dimensional}). Using the Wiener-Hopf solution~\eqref{eqn:V-0}, we approximate $\Ab_0(x)$ by using either $352$ effective wavenumbers $s_1, \, s_2, \, \ldots, s_{352}$, or just 1 effective wavenumber $s_1$. The Matching Method also accounts for multiple effective wavenumbers, and is described in~\cite{gower_multiple_2018}. The low volume fraction method assumes a low volume fraction expansion for just one effective wavenumber~\cite{linton_multiple_2005}. The small graph on the right is a magnification of the region around $x = 0$. Close to the boundary $x=0$, both $\Ab_0^1 \ee^{\ii s_1 x}$ and the low volume fraction method are inaccurate, which would potentially lead to inaccurate predictions for transmission and reflection.}
\label{fig:wiener-hopf-field}
\end{figure}

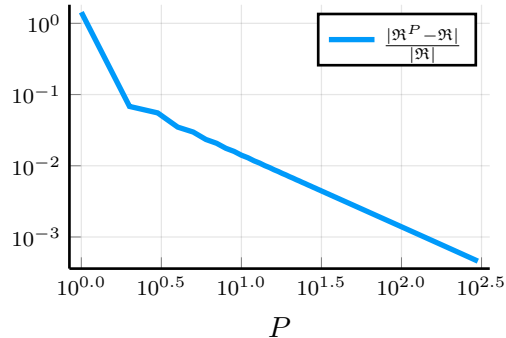
\begin{figure}[htb]
\centering
\input{"images/wiener-hopf-R.tex"}
\caption{ Demonstrates, with a log-log graph, how increasing the number of effective waves $P$ leads to a more accurate reflection coefficient $\mathfrak R^P$, when using~\eqref{eqn:R_P_0}. The non-dimensional wavenumber is $k b = 1.001$, and the other parameters used are given by~\eqref{eqn:parameters2}, with their definitions explained by (\ref{eqn:incident}--\ref{eqn:non-dimensional}). Here $\mathfrak R$ is the reflection coefficient given by~\eqref{eqn:D-0}. The error $|\mathfrak R^P - \mathfrak R|$ continuously drops as $P$ increases because of the rapid transition that occurs to $\Ab_0(x)$ near the boundary $x=0$, see~\Cref{fig:wiener-hopf-field}. However, methods such as the Matching Method~\cite{gower_multiple_2018} are able to accurately calculate the reflection coefficient without taking into account this rapid transition.}
\label{fig:wiener-hopf-R}
\end{figure}

For the numerical examples we use the parameters
\begin{equation}
  T_0 = - \frac{\mathrm J_0(k a_o)}{\mathrm H^{(1)}_0(k a_o)}, \quad  b = 1.001, \quad a_o =0.5, \quad \theta_\inc = \frac{\pi}{4}, \quad \phi = 30\%,
  \label{eqn:parameters2}
\end{equation}
which implies that the number fraction $\nfrac {} \approx 0.38$ per unit area. When we choose to fix the wavenumber, as we do for~\Cref{fig:wiener-hopf-field} and~\Cref{fig:wiener-hopf-R}, we use $b k = 1.001$. This leads to a wavelength ($2 \pi/k$) which is roughly six times larger than the particle diameter. If the particle was, say, more than a hundred times smaller than the wavelength, then only one effective wavenumber in the sum~\eqref{eqn:V-0} would be necessary to accurately calculate $\Ab_0(X)$.

To start we compare the average scattering coefficient $\Ab_0(x)$ calculated by the Wiener-Hopf solution~\eqref{eqn:V-0} with other methods in~\Cref{fig:wiener-hopf-field}. The most accurate of these other methods is the Matching Method~\cite{gower_multiple_2018,gower_effective_waves.jl:_2018}, and it closely agrees with the Wiener-Hopf solution  when using $352$ effective wavenumbers. The exception is the region close to the boundary $x=0$, where the Wiener-Hopf solution experiences a rapid transition. The low volume fraction method is the most commonly used in the literature: it assumes a small particle volume fraction\footnote{For the low volume fraction method we used a small volume fraction expansion for the wavenumber, but we numerically evaluated the wave amplitude. This is because the alternative, a small volume fraction expansion of the wave amplitude, led to poor results.} and just one effective wavenumber~\cite{linton_multiple_2005,martin_multiple_2011}. One significant conclusion we can draw from \Cref{fig:wiener-hopf-field} is that both the low volume fraction method and $\Ab_0^1 \ee^{\ii s_1 x}$ are inaccurate  near the boundary $x=0$. This means that both of these methods lead to inaccurate reflection coefficients.

In general, the Wiener-Hopf method does not lead to an explicit formula for the reflection coefficient~\eqref{eqn:R}, because we do not have an exact factorisation~\eqref{eqn:F-Split} for any truncated square matrices. However, there are methods~\cite{gower_multiple_2018,martin_multiple_2011,lawrie_brief_2007,abrahams_radiation_1996,abrahams_general_1990,abrahams_scattering_1987} to calculate $\Ab_n(x)$, from which we can obtain the reflection coefficient~\eqref{eqn:R-integral}. The method~\cite{gower_multiple_2018} also accounts for multiple effective wavenumbers. So one important question is: when using~\eqref{eqn:R-integral}, how many effective wavenumbers do we need to obtain an accurate reflection coefficient?

In~\Cref{fig:wiener-hopf-R} we show how increasing the number of effective waves $P$ reduces the error between $\mathfrak R^P$~\eqref{eqn:R_P_0} and $\mathfrak R$~\eqref{eqn:D-0}. To calculate a highly accurate reflection coefficient $\mathfrak R$, we could use either~\eqref{eqn:D-0} or the Matching Method~\cite{gower_multiple_2018,gower_effective_waves.jl:_2018}, as both give approximately the same $\mathfrak R$.

\begin{figure}[htb]
  \centering
  \resizebox{.99\linewidth}{!}{
    \includegraphics{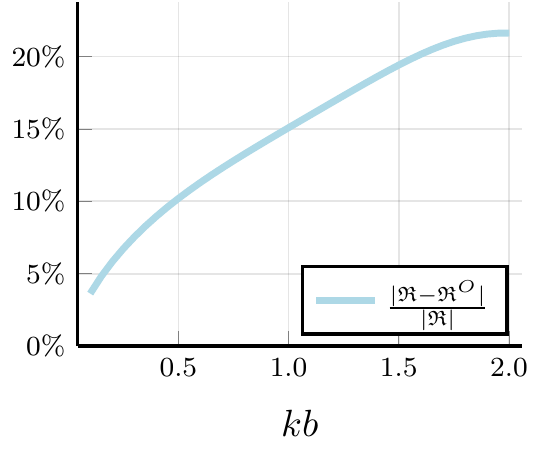}
    \includegraphics{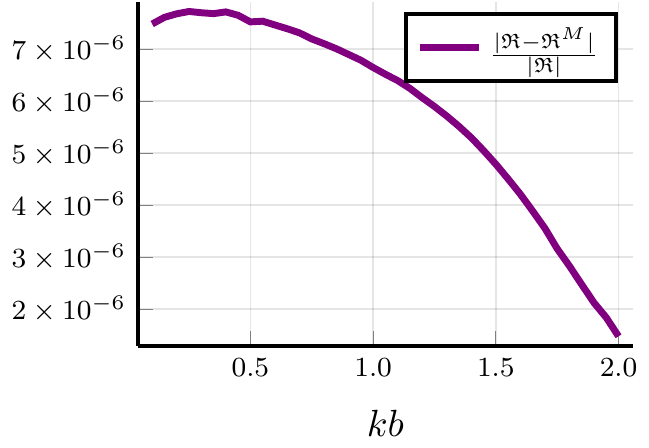}
  }
  \caption{Compares different methods for calculating the reflection coefficient when varying the non-dimensional wavenumber $k b$. The other parameters used are given by~\eqref{eqn:parameters2}, with their definitions explained in (\ref{eqn:incident}--\ref{eqn:non-dimensional}). Here $\mathfrak R$ is given by the Wiener-Hopf solution~\eqref{eqn:D-0}, $\mathfrak R^O$ uses a low volume fraction expansion of just one effective wavenumber~\cite{martin_multiple_2011}, and $\mathfrak R^M$ is calculated from the Matching Method~\cite{gower_multiple_2018}.}
  \label{fig:Reflection-Compare}
\end{figure}
Now we ask: how does the reflection coefficient~\eqref{eqn:R_P_0}, deduced via the Wiener-Hopf technique, compare with other methods across a broader range of wavenumbers. The result is shown in \Cref{fig:Reflection-Compare}, where $\mathfrak R^O$ is a low volume fraction expansion\footnote{We use the reflection coefficient~\cite[equation (39)]{martin_multiple_2011}, rather than the explicit low volume fraction expansion~\cite[equation (40 -- 41) ]{martin_multiple_2011}. This is because using equations (40 -- 41) led to roughly double the error we show.} of just one effective wavenumber~\cite{martin_multiple_2011}. The reflection coefficient $\mathfrak R^M$ is calculated from the Matching Method~\cite{gower_multiple_2018,gower_effective_waves.jl:_2018}. The general trend is clear: $\mathfrak R^O$ becomes more inaccurate as we increase the background wavenumber $k b$. On the other hand both $\mathfrak R^M$ and $\mathfrak R$ agree closely over all $k$.

One result to note is the ``instability`` exhibited by the Wiener-Hopf solution near the boundary $x=0$, see~\Cref{fig:wiener-hopf-field}. This instability occurs because we represented $\Ab_0(x)$ as a superposition of truncated waves, which is only accurate as long as the discarded terms are small. So, for a truncation number $P$, we can expect the instability to occur when $\ee^{\ii \s_P x}$ is not small, i.e.\ $x \approx 1/\mathrm{Im}\, s_p$. However, this instability does not affect the accuracy of the reflection coefficient~\eqref{eqn:D-0} deduced by the Wiener-Hopf technique, as demonstrated by close agreement with the Matching Method in~\Cref{fig:Reflection-Compare}.
%

\section{Conclusion and Next Steps}
\label{sec:Conclusion}
The major result of this paper is to prove that the ensemble-averaged field in random particulate materials consists of a superposition of waves, with complex effective wavenumbers, for one fixed incident wavenumber. These effective wavenumbers are governed by the dispersion equation~\eqref{eqn:detF} and are independent of the angle of incidence $\theta_\inc$. We showed asymptotically in ~\cref{sec:multiple-wavenumbers} that this has an infinite number of solutions, and hence there are an infinite number of effective wavenumbers. The Wiener-Hopf technique also provides a simple and elegant expression for the reflection coefficient~\eqref{eqn:R}, whose form can be used to guide and assess methods to characterise microstructure~\cite{roncen_bayesian_2018,gower_characterising_2018}.

To numerically implement the Wiener-Hopf technique, we considered particles that scatter only in their monopole mode in~\cref{sec:Numerical}. There we saw that when close to the interface of the half-space, a large number of effective wavenumbers were necessary to reach accurate agreement with an alternative method from the literature, the Matching Method as introduced by the authors in~\cite{gower_multiple_2018}. To obtain a constructive method via the Wiener-Hopf technique for general scatterers, and not just monopole scatterers, will require the factorisation of a matrix-function~\cite{rogosin_constructive_2016}, which is challenging. For these reasons the Matching Method~\cite{gower_multiple_2018} is presently more effective than using the Wiener-Hopf technique. However, there is ongoing work to use approximate methods~\cite{veitch_commutative_2007,abrahams_application_2000,abrahams_scattering_1987} which exploit the symmetry and properties of the matrix~\eqref{eqn:PsiToG}.

Moving forwards, this paper together with \cite{gower_multiple_2018}, establish accurate and robust solutions to the governing equation~\eqref{eqn:Wiener-Hopf}.
These same methods can now be translated to three spatial dimensions and vectorial waves (e.g.\ elasticity and electromagnetics), with much of the groundwork already available~\cite{linton_multiple_2006,kristensson_coherent_2015,conoir_effective_2010}.
Some clear challenges, that can now be addressed, are to verify the accuracy of the statistical assumptions used to deduce~\eqref{eqn:Wiener-Hopf}. These include the hole-correction and the quasicrystalline approximations. As these are now the only assumptions used, we could compare the solution of \eqref{eqn:Wiener-Hopf} with multipole methods~\cite{martin_multiple_2006,gower_multiplescattering.jl:_2018} in order to investigate their accuracy and limits of validity.


\appendix

\section{The Fourier transformed kernel $\hat \psi_n(s)$}
\label{app:hat_psi}
Here we calculate the Fourier transform~\eqref{eqn:FourierTransform} of $\psi_n(X)$~\eqref{eqn:integral_kernel}. To do so, it is simpler to use
\begin{equation}
  \mathrm F_n(X,Y) = (-1)^n \mathrm H^{(1)}_n(k R) \ee^{\ii n \Theta}.
  \label{eqn:Fn}
\end{equation}
Note that both $\mathrm F_n(X,Y)$ and $\ee^{\ii( {s} X + Y k \sin\theta_\inc)}$ satisfy wave equations, with
 \[
 \nabla^2 \mathrm F_n(X,Y) = - k^2 \mathrm F_n(X,Y) \quad \text{and} \quad
 \nabla^2 \ee^{\ii({s} X  + Y k\sin\theta_\inc)} = - S^2 \ee^{\ii( {s} X + Y k\sin\theta_\inc)},
 \]
where we used~\eqref{eqn:integral_kernel}${}_2$ for the first equation and~\eqref{eqn:SthetaS} for the second equation. This means that we can use Green's second identity to obtain
\begin{multline}
  (k^2 - S^2 ) \int_{\mathcal B} \ee^{\ii( {s} X + Y k \sin\theta_\inc)} \mathrm F_n(X,Y) \mathrm d X \mathrm d Y
  \\
   = \int_{\partial \mathcal B} \left [\frac{\partial \ee^{\ii( s X + Y K \sin\theta_\inc)}}{\partial \vec n} \mathrm F_n(X,Y) - \ee^{\ii( s X + K Y \sin\theta_\inc)} \frac{\partial \mathrm F_n(X,Y)}{\partial \vec n} \right] \mathrm d z,
   \label{eqn:Green2nd}
\end{multline}
for any area $\mathcal B$ in which the integrand is analytic,
where $\vec n$ is the outwards pointing unit normal and $\mathrm d z$ is a differential length along the boundary $\partial \mathcal B$. To calculate $\hat \psi_n({s})$, \rev{we take the region $\mathcal B$ to be defined by} $R \geq b$, with $(R,\Theta)$ being the polar coordinates of $(X,Y)$, in which case the integral over $\mathcal B$ converges because as $R \to \infty$ we have that
\begin{multline}
  |\ee^{\ii( {s} X + Y k \sin\theta_\inc)} \mathrm F_n(X,Y)| \sim
  \frac{|\ee^{ \ii {s} R \cos \Theta } \ee^{\ii k R (1 + \sin\Theta \sin\theta_\inc) }| }{\sqrt{ \pi |k| R/2}}
  \\
  \leq \frac{ |\ee^{ - R ( \mathrm{Im} \, k (1 - |\sin \theta_\inc|) - |\mathrm{Im}\,({s})|} )| }{\sqrt{ \pi |k| R/2}} \to 0,
\end{multline}
 exponentially fast when $|\mathrm{Im}\,({s})| < \mathrm{Im}\, k (1 - |\sin \theta_\inc|)$. Under this restriction, and by assuming $S \not = \pm k$, \eqref{eqn:Green2nd} then leads to
\begin{equation}
 \hat \psi_n({s}) = \int_{R \geq b} \ee^{\ii {s} X + \ii k Y \sin\theta_\inc} \mathrm F_n(X,Y) \mathrm d X \mathrm d Y
= \frac{\mathcal I_n(b)}{k^2 - S^2},
 \label{eqn:KernalFourier}
\end{equation}
by using ${s} X + Y k \sin\theta_\inc = R S \cos(\theta - \theta_S)$ from \eqref{eqn:SthetaS} and
\begin{multline}
  \mathcal I_n (R) =  \int_{0}^{2 \pi} -\frac{\partial \ee^{ \ii S R  \cos(\Theta - \theta_S)}}{\partial R} \mathrm F_n(k \vec X)
    + \ee^{\ii S R  \cos(\Theta - \theta_S)} \frac{\partial \mathrm F_n(k \vec X)}{\partial R}  R \mathrm d \Theta
  \\
  =
(-1)^n\int_{0}^{2 \pi} \ee^{\ii S R \cos(\Theta - \theta_S)} \ee^{\ii n \Theta} \big[k \mathrm H^{(1)\prime}_n(k R) - \ii S  \cos(\Theta - \theta_S) \mathrm H^{(1)}_n(k R) \big ]  R \mathrm d \theta
   \\
  =
 (-1)^n\int_{0}^{2 \pi} \sum_{m=-\infty}^\infty \ii^m \mathrm J_m(S R)
  \ee^{\ii m (\Theta - \theta_S)}  \Big [ k \ee^{\ii n \Theta} \mathrm H^{(1)\prime}_n(k R) \hspace{3cm}
  \\
   - \frac{\ii S}{2} (\ee^{\ii (n+1) \Theta - \ii \theta_S} + \ee^{\ii (n-1) \Theta + \ii \theta_S} ) \mathrm H^{(1)}_n(k R) \Big ]  R \mathrm d \Theta
   \\
  =
  2 \pi (-\ii)^{n} R \ee^{\ii n \theta_S} \left [
  k \mathrm J_{n}(S R) \mathrm H^{(1)\prime}_n(k R) - S \mathrm J_n'(S R) \mathrm H^{(1)}_n(k R)
  \right ],
   \label{eqn:In}
\end{multline}
where $\mathrm J_n$ is the Bessel function of the first kind, and we used the Jacobi-Anger expansion on $\ee^{\ii S R \cos(\Theta - \theta_S)}$, integrated over $\Theta$ and used the identity $\mathrm J_{n-1}(SR) - \mathrm J_{n+1}(SR) = 2 \mathrm J_n'(SR)$. In summary
\begin{equation}
  \hat \psi_n(s) =  2 \pi \frac{(-\ii)^{n} \ee^{\ii n \theta_S}}{{\alpha}^2 - {s}^2}  \mathrm N_{n}(b S),
  \label{eqn:KernalFourierResult}
\end{equation}
when the condition~\eqref{eqn:ImsImK} is satisfied, with $\mathrm N_n$ given by~\eqref{eqn:N_comp}.

Below we establish some useful properties for $\hat \psi_n(s)$. In particular, we show that $\hat \psi_n(s)$ has no branch-points.


The function $\mathrm N_{n}(b S)$, for integer values of $n$, can be expanded around $S =0$ as
\begin{equation}
  \mathrm N_{n}(b S) = S^{|n|} \sum_{m=0}^\infty c_{m|n|} S^{2m},
  \label{eqn:N-expand}
\end{equation}
where the $c_{m|n|}$ are some constants that depend on $m$ and $|n|$, and the radius of convergence of the series above is infinite. Using~\eqref{eqn:SthetaS} we can write
\begin{equation}
  \ee^{\ii n \theta_S } = \ee^{\ii \sgn(n) |n| \theta_S } = (\cos \theta_S + \sgn(n) \ii \sin \theta_S)^{|n|} = (s + \sgn(n) \ii k \sin \theta_\inc)^{|n|} S^{-|n|}.
  \label{eqn:theta-S-expand}
\end{equation}

Substituting~\eqref{eqn:N-expand} and~\eqref{eqn:theta-S-expand} in \eqref{eqn:KernalFourierResult} results in
\begin{equation}
  \hat \psi_n(s) = \frac{ 2 \pi (-\ii)^{n}}{{\alpha}^2 - {s}^2}  (s + \sgn(n) \ii k \sin \theta_\inc)^{|n|} \sum_{m=0}^\infty c_{m|n|} S^{2m}.
  \label{eqn:KernalFourierResult-expand}
\end{equation}
Because $S^2 = s^2 + k^2 \sin^2 \theta_\inc$, we can immediately see from the above that $\hat \psi_n(s)$ has no branch-points. Additionally we can establish the properties:
\begin{equation}
  \hat \psi_n(s) = \hat \psi_{-n}(-s) = \hat \psi_n(-s) \ee^{2 \ii n \theta_S}(-1)^n.
  \label{eqn:psi-properties}
\end{equation}

\section{Asymptotic location of the wavenumbers}
\label{app:asymptotic-wavenumbers}
Here we explicitly calculate a sequence of effective wavenumbers $S_p$, assuming $|S_p|$ large and increasing with $p$, and $\mathrm{Im}\, S_p >0$, that asymptotically satisfy~\eqref{eqn:detF}.
A key step is to approximate the terms appearing in~\eqref{eqn:F_comp}, such as
\begin{equation}
\mathrm J_n(b S) \sim \frac{\ee^{\frac{\ii \pi }{4} + \frac{\ii n \pi}{2} - \ii b S}}{\sqrt{2 \pi b S}}
\quad \text{and} \quad
\mathrm J_n'(b S) \sim \frac{\ee^{-\frac{\ii \pi }{4} + \frac{\ii n \pi}{2} - \ii b S}}{\sqrt{2 \pi b S}},
\label{eqns:Jasym}
\end{equation}
for large $|b S|$, where the terms $\ee^{\ii b S}$ are discarded as Im $b S \to \infty$.

\paragraph{Monopole scatterers} The simplest case is for monopole scatterers, where $n=m =0$ in~\eqref{eqn:F_comp}, and the effective wavenumber $S$ satisfies
\begin{equation}
  b^2 \det \mathbf G = (b S)^2 - (b k)^2 + 2 \pi \nfrac {} b^2 T_0 \mathrm N_0(bS) \sim
  (b S)^2 - c \sqrt{b S} \ee^{- \ii b S} = 0,
 \label{eqn:dispersion0}
\end{equation}
 where $c = \sqrt{2\pi} \nfrac {} b^2 T_0 \mathrm H^{(1)}_0(k b)\ee^{-\frac{\ii \pi }{4}}$. Here we used~\eqref{eqns:Jasym}, and ignored terms  which are algebraically smaller than $bS$. To find the root of the above we substitute
\begin{equation}
  b S = x + \ii \log y,
  \label{eqn:Sasym}
\end{equation}
where $x$ and $y$ are real, and $|x|$ and $y$ are large with $y > 1$. This leads to
\begin{equation}
  (x+ \ii \log y)^{3/2} - c \ee^{- \ii x} y = 0.
\end{equation}
For the logarithm and square root we use the typical branch cut $(-\infty,0)$ and take positive values of the functions for positive arguments.
For the above to be satisfied to leading order then $x^{3/2} \sim y$, which reduces the above equation to
\begin{equation}
  x^{3/2} \sim r_c \ee^{\ii (\theta_c -x)} y,
\end{equation}
where we substituted $c = r_c \ee^{\ii \theta_c}$, for real scalars $r_c$ and $\theta_c$.
Equating the real and imaginary parts of the above leads to
\begin{align}
& x \sim \theta_c + 2\pi p \quad \text{and} \quad y \sim \frac{1}{r_c}(\theta_c + 2\pi p)^{3/2} \quad
\text{for} \;\;  p > -\frac{\theta_c}{2 \pi}
\\
& x \sim \theta_c - \frac{3 \pi}{2} + 2\pi p \quad \text{and} \quad y \sim \frac{1}{r_c}(-\theta_c + \frac{3 \pi}{2} - 2\pi p)^{3/2} \quad
\text{for} \;\;   p < \frac{3}{4} - \frac{\theta_c}{2\pi},
\label{eqns:monopole-asym}
\end{align}
for integers $p$. From this we can identify that, at leading order, the effective wavenumbers are given by~\eqref{eqn:asymS_p}.

\paragraph{Multipole scatterers:} With the same method used above, we can also demonstrate the existence of multiple effective wavenumbers for $n, m = -M,-M+1,\cdots,M$ in~\eqref{eqn:F_comp}. To show this explicitly, we consider $b k$ to be the same order as $b S$, that is $|k| \sim |S|$.

By considering $b k$ large, we can approximate
\begin{equation}
  \mathrm H^{(1)}_n(b k)  \sim  \ee^{\ii\left( b k - \frac{\pi}{4} - \frac{n \pi}{2} \right)} \sqrt{\frac{2}{\pi b k}} \quad \text{and} \quad
  \mathrm H^{(1)\prime}_n(b k) \sim  \ee^{\ii (b k +\frac{\pi}{4} - \frac{n \pi}{2} )} \sqrt{\frac{2}{\pi b k}},
\end{equation}
combining this with~\eqref{eqns:Jasym} and considering $|k| \sim |S|$, the term~\eqref{eqn:F_comp} at leading order becomes
\begin{equation}
  b^2 G_{m n} = d_0 \delta_{m n} +  c_0 T_m,
\end{equation}
where
\[
d_0 = (b S)^2 - (b k)^2, \quad \text{and} \quad c_0 =  2 \nfrac{} b^2 \frac{\ii (k+S)}{\sqrt{k S}} \ee^{\ii b(k- S)}.
\]
By simple rearrangement of the determinant we find that\footnote{The determinant of $b^2 \vec G$ equals the product of its eigenvalues. The eigenvector $(T_{-M},\cdots,T_M)^\mathrm T$ gives the eigenvalue $d_0 + c_0 \sum_m T_m$, while all other eigenvalues equal $d_0$.}
\begin{equation}
  \det(b^2 \mathbf G) = d_0^{2 M} \left (d_0 + c_0\sum_{m=-M}^M T_m \right).
\end{equation}
Note that $d_0 \not= 0$, i.e.\ $S \not= \pm k$, was necessary to reach the condition~\eqref{eqn:ImsImK}, which was used to calculate the Fourier transforms~\eqref{eqn:WH-Fourier-Matrix}. Taking this into consideration, and taking the limit $M \to \infty$, the effective wavenumbers $S$ must satisfy
\begin{equation}
  d_0 + c_0\sum_{m=-M}^M T_m = 0 \implies b S - b k  = - 2 \nfrac{} \ii b^2 \sum_{m=-\infty}^\infty  T_m \frac{\ee^{\ii b(k- S)}}{b\sqrt{k S}}  .
  \label{eqn:pre-SMasym}
\end{equation}
Using an asymptotic expansion analogous to~\eqref{eqn:Sasym}, the above leads to the effective wavenumbers~\eqref{eqn:SMasym}.

\section{Equivalent determinants}
\label{app:formulas}
For any square matrices $\mathbf A$ and $\vec B$, and scalar $c$, if $A_{nm} = B_{nm} c^{n-m}$ (not employing the summation convention), then
\begin{equation}
  \det \mathbf A = \det \vec B.
  \label{eqn:DetAtoB}
\end{equation}
This follows simply by defining the diagonal matrix $C_{nm} = \delta_{nm} c^n$, which leads to $\vec A = \vec C \vec B \vec C^{-1}$, and $\det (\vec C \vec B \vec C^{-1}) = \det \vec C  \det \vec B \det \vec C^{-1} = \det \vec B$.

\dataccess{We provide the code to generate all the graphs in~\cite{gower_effective_waves.jl:_2018}.}

\aucontribute{ I.D.A and A.L.G. conceived of the study. A.L.G. drafted the manuscript. A.L.G., I.D.A, and W.J.P edited the manuscript and gave final approval for publication.}

\competing{We have no competing interests.}

\funding{
This work was funded by EPSRC (EP/M026205/1,EP/L018039/1,EP/S019804/1) including via the Isaac Newton Institute (EP/K032208/1,EP/R014604/1) and partial support from the UKAN grant EPSRC (EP/R005001/1).
}

\bibliographystyle{RS}
\bibliography{library,SharedMultipleScattering}

\end{document}

%% file: preamble.tex
\usepackage[utf8]{inputenc} 
\usepackage[T1]{fontenc}

\usepackage{hyperref}
\usepackage{cleveref}

\usepackage{amsmath,amssymb,mathtools}
\usepackage{verbatim}

\usepackage{esint} 

\usepackage{lipsum}
\usepackage{amsfonts}

\usepackage{graphicx,psfrag,color,xcolor}
\usepackage{tikz,pgfplots}
\usetikzlibrary{patterns}

\usepackage{epstopdf}
\ifpdf
  \DeclareGraphicsExtensions{.eps,.pdf,.png,.jpg}
\else
  \DeclareGraphicsExtensions{.eps}
\fi

\usepackage{amsopn}

%% file: macros.tex
\def\rev#1{\textcolor{black}{#1}}	    

\newcommand{\ii}{\textrm{i}}
\newcommand{\ee}{\textrm{e}}

\newcommand{\been}{\begin{equation}}
\newcommand{\een}{\end{equation}}
\newcommand{\beena}{\begin{eqnarray}}
\newcommand{\eena}{\end{eqnarray}}

\renewcommand{\vec}[1]{\bm{#1}}

\def\XXint#1#2#3{{\setbox0=\hbox{$#1{#2#3}{\int}$}
     \vcenter{\hbox{$#2#3$}}\kern-.5\wd0}}

\DeclareMathOperator{\sgn}{sgn}


%% file: mst-macros.tex
\usepackage[colorinlistoftodos,bordercolor=orange,backgroundcolor=orange!20,linecolor=orange,textsize=scriptsize]{todonotes}
\usepackage{soul}

\newcommand{\edit}[1]{\textcolor{black}{#1}}

\newcommand \s {\mathbf s}
\newcommand \Ab {A}
\newcommand \Db {D}

\newcommand {\nfrac}[1] {\mathfrak n_{#1}}

\newcommand \inc{\mathrm{inc}}

\newcommand{\ensem}[1]{\langle #1 \rangle}

%% file: images/multiple-waves-diagram.tex
\begin{tikzpicture}[]
\begin{axis}[height = {76.82000000000001mm}, axis equal = {true}, ylabel = {}, xmin = {-0.9852330780522804}, xmax = {3.1885627445369504}, ymax = {1.6853921066052227}, xlabel = {}, unbounded coords=jump,scaled x ticks = false,xlabel style = {font = {\fontsize{11 pt}{14.3 pt}\selectfont}, color = {rgb,1:red,0.00000000;green,0.00000000;blue,0.00000000}, draw opacity = 1.0, rotate = 0.0},xmajorticks=false,xmajorgrids = false,axis lines* = left,separate axis lines,x axis line style = {draw opacity = 0},scaled y ticks = false,ylabel style = {font = {\fontsize{11 pt}{14.3 pt}\selectfont}, color = {rgb,1:red,0.00000000;green,0.00000000;blue,0.00000000}, draw opacity = 1.0, rotate = 0.0},ymajorticks=false,ymajorgrids = false,axis lines* = left,separate axis lines,y axis line style = {draw opacity = 0},    xshift = 0.0mm,
    yshift = 0.0mm,
    axis background/.style={fill={rgb,1:red,1.00000000;green,1.00000000;blue,1.00000000}}
,legend style = {color = {rgb,1:red,0.00000000;green,0.00000000;blue,0.00000000},
draw opacity = 1.0,
line width = 1,
solid,fill = {rgb,1:red,1.00000000;green,1.00000000;blue,1.00000000},font = {\fontsize{8 pt}{10.4 pt}\selectfont}},colorbar style={title=}, ymin = {-1.4853921066052227}, width = {125.34799999999998mm}]\addplot+ [color = {rgb,1:red,0.00000000;green,0.00000000;blue,0.00000000},
draw opacity = 0.0,
line width = 2.0,
solid,mark = none,
mark size = 2.0,
mark options = {
    color = {rgb,1:red,0.00000000;green,0.00000000;blue,0.00000000}, draw opacity = 0.5,
    fill = {rgb,1:red,0.00000000;green,0.60560316;blue,0.97868012}, fill opacity = 0.5,
    line width = 1,
    rotate = 0,
    solid
},fill = {rgb,1:red,0.67843137;green,0.84705882;blue,0.90196078}, fill opacity=1.0,forget plot]coordinates {
(0.0, -1.3956529307596441)
(3.0704364476712174, -1.3956529307596441)
(3.0704364476712174, 1.595652930759644)
(0.0, 1.595652930759644)
};
\addplot+ [color = {rgb,1:red,0.00000000;green,0.00000000;blue,0.00000000},
draw opacity = 1.0,
line width = 2.0,
solid,mark = none,
mark size = 2.0,
mark options = {
    color = {rgb,1:red,0.00000000;green,0.00000000;blue,0.00000000}, draw opacity = 1.0,
    fill = {rgb,1:red,0.00000000;green,0.00000000;blue,0.00000000}, fill opacity = 1.0,
    line width = 1,
    rotate = 0,
    solid
},fill = {rgb,1:red,0.00000000;green,0.00000000;blue,0.00000000}, fill opacity=1.0,forget plot]coordinates {
(0.4, -1.1956529307596442)
(0.5644860787516464, -1.1225480068700238)
(0.5604246940911118, -1.1134098913838213)
(0.5827623097240515, -1.1144252375489547)
(0.5685474634121809, -1.1316861223562262)
(0.5644860787516464, -1.1225480068700238)
};
\addplot+ [color = {rgb,1:red,0.00000000;green,0.00000000;blue,0.00000000},
draw opacity = 1.0,
line width = 2.0,
solid,mark = none,
mark size = 2.0,
mark options = {
    color = {rgb,1:red,0.00000000;green,0.00000000;blue,0.00000000}, draw opacity = 1.0,
    fill = {rgb,1:red,0.00000000;green,0.00000000;blue,0.00000000}, fill opacity = 1.0,
    line width = 1,
    rotate = 0,
    solid
},fill = {rgb,1:red,0.00000000;green,0.00000000;blue,0.00000000}, fill opacity=1.0,forget plot]coordinates {
(0.7181980515339463, -0.4325908223786446)
(0.4318198051533947, -0.7189690687591963)
(0.447729707730092, -0.7348789713358936)
(0.4, -0.750788873912591)
(0.41590990257669735, -0.703059166182499)
(0.4318198051533947, -0.7189690687591963)
};
\addplot+ [color = {rgb,1:red,0.00000000;green,0.00000000;blue,0.00000000},
draw opacity = 1.0,
line width = 2.0,
solid,mark = none,
mark size = 2.0,
mark options = {
    color = {rgb,1:red,0.00000000;green,0.00000000;blue,0.00000000}, draw opacity = 1.0,
    fill = {rgb,1:red,0.00000000;green,0.00000000;blue,0.00000000}, fill opacity = 1.0,
    line width = 1,
    rotate = 0,
    solid
},fill = {rgb,1:red,0.00000000;green,0.00000000;blue,0.00000000}, fill opacity=1.0,forget plot]coordinates {
(0.4, 0.6739026841148619)
(0.8269074841227312, 0.8162051788224389)
(0.8190017899723102, 0.8399222612737017)
(0.8743416490252569, 0.8320165671232809)
(0.8348131782731522, 0.7924880963711761)
(0.8269074841227312, 0.8162051788224389)
};
\addplot+ [color = {rgb,1:red,0.00000000;green,0.00000000;blue,0.00000000},
draw opacity = 1.0,
line width = 2.0,
solid,mark = none,
mark size = 2.0,
mark options = {
    color = {rgb,1:red,0.00000000;green,0.00000000;blue,0.00000000}, draw opacity = 1.0,
    fill = {rgb,1:red,0.00000000;green,0.00000000;blue,0.00000000}, fill opacity = 1.0,
    line width = 1,
    rotate = 0,
    solid
},fill = {rgb,1:red,0.00000000;green,0.00000000;blue,0.00000000}, fill opacity=1.0,forget plot]coordinates {
(0.4, 1.1956529307596442)
(0.5739222891282244, 1.242032207860504)
(0.5713456626226211, 1.2516945572565163)
(0.5932469879202493, 1.2471854608717106)
(0.5764989156338277, 1.2323698584644915)
(0.5739222891282244, 1.242032207860504)
};
\addplot+ [color = {rgb,1:red,0.00000000;green,0.00000000;blue,0.00000000},
draw opacity = 1.0,
line width = 2.0,
solid,mark = none,
mark size = 2.0,
mark options = {
    color = {rgb,1:red,0.00000000;green,0.00000000;blue,0.00000000}, draw opacity = 1.0,
    fill = {rgb,1:red,0.00000000;green,0.00000000;blue,0.00000000}, fill opacity = 1.0,
    line width = 1,
    rotate = 0,
    solid
},fill = {rgb,1:red,0.00000000;green,0.00000000;blue,0.00000000}, fill opacity=1.0,forget plot]coordinates {
(-0.8671067811865475, -0.7071067811865475)
(-0.23071067811865478, -0.07071067811865475)
(-0.2660660171779822, -0.035355339059327376)
(-0.16000000000000003, 0.0)
(-0.1953553390593274, -0.10606601717798213)
(-0.23071067811865478, -0.07071067811865475)
};
\addplot+ [color = {rgb,1:red,0.00000000;green,0.00000000;blue,0.00000000},
draw opacity = 1.0,
line width = 2.0,
dashed,mark = none,
mark size = 2.0,
mark options = {
    color = {rgb,1:red,0.00000000;green,0.00000000;blue,0.00000000}, draw opacity = 1.0,
    fill = {rgb,1:red,0.00000000;green,0.00000000;blue,0.00000000}, fill opacity = 1.0,
    line width = 1,
    rotate = 0,
    solid
},forget plot]coordinates {
(-0.06, -1.2246467991473533e-17)
(-0.86, 8.572527594031472e-17)
};
\addplot+ [color = {rgb,1:red,0.00000000;green,0.00000000;blue,0.00000000},
draw opacity = 1.0,
line width = 2.0,
dotted,mark = none,
mark size = 2.0,
mark options = {
    color = {rgb,1:red,0.00000000;green,0.00000000;blue,0.00000000}, draw opacity = 1.0,
    fill = {rgb,1:red,0.00000000;green,0.00000000;blue,0.00000000}, fill opacity = 1.0,
    line width = 1,
    rotate = 0,
    solid
},forget plot]coordinates {
(-0.51, 4.286263797015736e-17)
(-0.5099972740926235, -0.0013813499675771677)
(-0.5099890964129546, -0.0027626784183995595)
(-0.5099754670883739, -0.004143963836047939)
(-0.50995638633118, -0.005525184704772956)
(-0.5099318544385866, -0.006906319509830716)
(-0.5099018717927173, -0.00828734673781837)
(-0.5098664388606003, -0.00966824487700829)
(-0.5098255561941608, -0.011048992417683612)
(-0.5097792244302122, -0.012429567852473754)
(-0.5097274442904465, -0.013809949676688344)
(-0.5096702165814229, -0.015190116388653422)
(-0.5096075421945553, -0.016570046490045284)
(-0.5095394221060984, -0.01794971848622582)
(-0.5094658573771326, -0.01932911088657779)
(-0.5093868491535471, -0.020708202204838648)
(-0.5093023986660227, -0.022086970959435674)
(-0.5092125072300118, -0.023465395673821075)
(-0.5091171762457185, -0.024843454876805574)
(-0.5090164071980771, -0.026221127102893326)
(-0.5089102016567276, -0.02759839089261672)
(-0.508798561275993, -0.02897522479286977)
(-0.5086814877948523, -0.03035160735724269)
(-0.5085589830369139, -0.03172751714635644)
(-0.508431048910387, -0.03310293272819575)
(-0.5082976874080523, -0.03447783267844346)
(-0.5081589006072299, -0.03585219558081463)
(-0.5080146906697486, -0.03722600002738925)
(-0.5078650598419108, -0.03859922461894619)
(-0.5077100104544578, -0.03997184796529692)
(-0.5075495449225343, -0.041343848685617676)
(-0.5073836657456499, -0.042715205408783714)
(-0.5072123755076402, -0.04408589677370113)
(-0.5070356768766272, -0.045455901429640046)
(-0.5068535726049772, -0.046825198036567636)
(-0.5066660655292581, -0.048193765265479624)
(-0.506473158570195, -0.04956158179873296)
(-0.5062748547326252, -0.0509286263303784)
(-0.5060711571054509, -0.052294877566491337)
(-0.5058620688615914, -0.05366031422550408)
(-0.5056475932579333, -0.05502491503853773)
(-0.5054277336352804, -0.05638865874973258)
(-0.5052024934183009, -0.057751524116579654)
(-0.5049718761154748, -0.0591134899102521)
(-0.5047358853190387, -0.060474534915934866)
(-0.5044945247049298, -0.06183463793315568)
(-0.5042477980327295, -0.06319377777611575)
(-0.5039957091456035, -0.0645519332740186)
(-0.5037382619702431, -0.06590908327140119)
(-0.5034754605168036, -0.06726520662846228)
(-0.5032073088788415, -0.0686202822213922)
(-0.5029338112332512, -0.0699742889427024)
(-0.5026549718402, -0.07132720570155324)
(-0.5023707950430615, -0.07267901142408299)
(-0.5020812852683475, -0.07402968505373661)
(-0.5017864470256401, -0.07537920555159272)
(-0.5014862849075208, -0.07672755189669188)
(-0.5011808035894987, -0.07807470308636437)
(-0.5008700078299382, -0.07942063813655656)
(-0.500553902469985, -0.0807653360821581)
(-0.5002324924334899, -0.08210877597732902)
(-0.4999057827269331, -0.08345093689582504)
(-0.49957377843934525, -0.08479179793132403)
(-0.4992364847422288, -0.0861313381977521)
(-0.49889390688947755, -0.0874695368296078)
(-0.49854605021729415, -0.08880637298228848)
(-0.49819292014410765, -0.09014182583241383)
(-0.49783452217048874, -0.09147587457815064)
(-0.4974708618790641, -0.09280849843953747)
(-0.4971019449344296, -0.09413967665880721)
(-0.49672777708306204, -0.09546938850071097)
(-0.49634836415322925, -0.09679761325284153)
(-0.4959637120549, -0.09812433022595503)
(-0.4955738267796512, -0.09944951875429371)
(-0.49517871440057537, -0.10077315819590821)
(-0.4947783810721851, -0.1020952279329783)
(-0.4943728330303182, -0.10341570737213443)
(-0.4939620765920396, -0.1047345759447789)
(-0.4935461181555437, -0.10605181310740547)
(-0.49312496420005436, -0.10736739834191969)
(-0.49269862128572395, -0.10868131115595907)
(-0.49226709605353136, -0.10999353108321122)
(-0.49183039522517835, -0.11130403768373331)
(-0.49138852560298485, -0.11261281054427069)
(-0.4909414940697835, -0.11391982927857398)
(-0.4904893075888115, -0.11522507352771769)
(-0.4900319732036029, -0.11652852296041638)
(-0.48956949803787897, -0.11783015727334177)
(-0.4891018892954364, -0.11912995619143944)
(-0.48862915426003606, -0.12042789946824377)
(-0.48815130029528875, -0.12172396688619372)
(-0.4876683348445412, -0.1230181382569483)
(-0.48718026543075943, -0.12431039342169993)
(-0.4866870996564122, -0.1256007122514892)
(-0.48618884520335215, -0.1268890746475185)
(-0.4856855098326964, -0.12817546054146461)
(-0.4851771013847057, -0.12945984989579132)
(-0.4846636277786617, -0.13074222270406224)
(-0.48414509701274444, -0.13202255899125162)
(-0.48362151716390733, -0.13330083881405563)
(-0.4830928963877512, -0.13457704226120376)
(-0.4825592429183979, -0.13585114945376775)
(-0.4820205650683611, -0.13712314054547253)
(-0.48147687122841754, -0.13839299572300418)
(-0.4809281698674761, -0.1396606952063192)
(-0.4803744695324458, -0.1409262192489529)
(-0.4798157788481029, -0.14218954813832613)
(-0.47925210651695616, -0.14345066219605282)
(-0.47868346131911177, -0.14470954177824688)
(-0.4781098521121362, -0.14596616727582737)
(-0.47753128783091847, -0.1472205191148244)
(-0.4769477774875308, -0.14847257775668432)
(-0.4763593301710882, -0.14972232369857336)
(-0.4757659550476073, -0.15096973747368178)
(-0.4751676613598631, -0.1522147996515275)
(-0.47456445842724504, -0.15345749083825796)
(-0.4739563556456122, -0.1546977916769526)
(-0.4733433624871467, -0.15593568284792483)
(-0.4727254885002059, -0.1571711450690219)
(-0.47210274330917434, -0.15840415909592656)
(-0.47147513661431306, -0.15963470572245556)
(-0.4708426781916092, -0.1608627657808595)
(-0.47020537789262307, -0.16208832014212166)
(-0.46956324564433516, -0.16331134971625522)
(-0.4689162914489915, -0.16453183545260094)
(-0.46826452538394736, -0.16574975834012445)
(-0.4676079576015111, -0.16696509940771134)
(-0.4669465983287854, -0.16817783972446337)
(-0.4662804578675078, -0.16938796039999351)
(-0.4656095465938911, -0.17059544258471945)
(-0.46493387495846095, -0.1718002674701577)
(-0.4642534534858931, -0.17300241628921684)
(-0.4635682927748501, -0.1742018703164891)
(-0.4628784034978156, -0.17539861086854247)
(-0.4621837964009282, -0.17659261930421208)
(-0.4614844823038142, -0.1777838770248898)
(-0.4607804720994192, -0.17897236547481427)
(-0.4600717767538379, -0.18015806614136043)
(-0.4593584073061442, -0.18134096055532697)
(-0.45864037486821796, -0.18252103029122493)
(-0.4579176906245731, -0.1836982569675639)
(-0.457190365832183, -0.18487262224713868)
(-0.45645841182030467, -0.18604410783731545)
(-0.4557218399903029, -0.18721269549031566)
(-0.4549806618154728, -0.1883783670035009)
(-0.45423488884085994, -0.1895411042196567)
(-0.4534845326830821, -0.19070088902727467)
(-0.45272960503014703, -0.1918577033608348)
(-0.45197011764127115, -0.19301152920108755)
(-0.4512060823466959, -0.1941623485753336)
(-0.45043751104750407, -0.19531014355770396)
(-0.4496644157154335, -0.19645489626944015)
(-0.44888680839269157, -0.1975965888791713)
(-0.44810470119176726, -0.19873520360319274)
(-0.44731810629524194, -0.19987072270574316)
(-0.4465270359556004, -0.20100312849927993)
(-0.4457315024950397, -0.20213240334475596)
(-0.4449315183052771, -0.20325852965189314)
(-0.4441270958473573, -0.20438148987945706)
(-0.443318247651458, -0.20550126653553047)
(-0.4425049863166949, -0.20661784217778492)
(-0.44168732451092574, -0.207731199413753)
(-0.4408652749705523, -0.20884132090109953)
(-0.44003885050032276, -0.20994818934789092)
(-0.43920806397313183, -0.211051787512865)
(-0.43837292832982, -0.21215209820569986)
(-0.43753345657897247, -0.21324910428728086)
(-0.43668966179671653, -0.21434278866996803)
(-0.4358415571265173, -0.21543313431786268)
(-0.43498915577897346, -0.21652012424707184)
(-0.4341324710316117, -0.21760374152597325)
(-0.4332715162286789, -0.2186839692754796)
(-0.43240630478093556, -0.21976079066930032)
(-0.43153685016544596, -0.22083418893420498)
(-0.4306631659253687, -0.22190414735028352)
(-0.4297852656697456, -0.22297064925120688)
(-0.4289031630732898, -0.22403367802448743)
(-0.4280168718761723, -0.2250932171117366)
(-0.427126405883809, -0.2261492500089233)
(-0.4262317789666442, -0.22720176026663144)
(-0.425333005059936, -0.22825073149031527)
(-0.42443009816353794, -0.22929614734055537)
(-0.42352307234168196, -0.23033799153331314)
(-0.42261194172275884, -0.23137624784018407)
(-0.42169672049909823, -0.23241090008865084)
(-0.42077742292674725, -0.23344193216233536)
(-0.4198540633252491, -0.23446932800124928)
(-0.4189266560774195, -0.23549307160204463)
(-0.4179952156291227, -0.23651314701826304)
(-0.4170597564890466, -0.23752953836058383)
(-0.4161202932284769, -0.23854222979707176)
(-0.4151768404810696, -0.23955120555342382)
(-0.41422941294262383, -0.2405564499132144)
(-0.413278025370852, -0.2415579472181407)
(-0.41232269258515075, -0.2425556818682661)
(-0.41136342946636983, -0.24354963832226337)
(-0.41040025095657995, -0.2445398010976572)
(-0.4094331720588408, -0.24552615477106454)
(-0.4084622078369666, -0.24650868397843517)
(-0.4074873734152917, -0.2474873734152916)
};
\addplot+ [color = {rgb,1:red,0.00000000;green,0.00000000;blue,0.00000000},
draw opacity = 1.0,
line width = 2.0,
solid,mark = none,
mark size = 2.0,
mark options = {
    color = {rgb,1:red,0.00000000;green,0.00000000;blue,0.00000000}, draw opacity = 1.0,
    fill = {rgb,1:red,0.00000000;green,0.00000000;blue,0.00000000}, fill opacity = 1.0,
    line width = 1,
    rotate = 0,
    solid
},fill = {rgb,1:red,0.00000000;green,0.00000000;blue,0.00000000}, fill opacity=1.0,forget plot]coordinates {
(-0.16, -0.0)
(-0.6691168824543142, 0.5091168824543142)
(-0.6974011537017761, 0.4808326112068523)
(-0.725685424949238, 0.565685424949238)
(-0.6408326112068523, 0.5374011537017761)
(-0.6691168824543142, 0.5091168824543142)
};
\addplot+ [color = {rgb,1:red,0.00000000;green,0.00000000;blue,0.00000000},
draw opacity = 1.0,
line width = 2.0,
dashed,mark = none,
mark size = 2.0,
mark options = {
    color = {rgb,1:red,0.00000000;green,0.00000000;blue,0.00000000}, draw opacity = 1.0,
    fill = {rgb,1:red,0.00000000;green,0.00000000;blue,0.00000000}, fill opacity = 1.0,
    line width = 1,
    rotate = 0,
    solid
},forget plot]coordinates {
(-0.06, -1.2246467991473533e-17)
(-0.86, 8.572527594031472e-17)
};
\addplot+ [color = {rgb,1:red,0.00000000;green,0.00000000;blue,0.00000000},
draw opacity = 1.0,
line width = 2.0,
dotted,mark = none,
mark size = 2.0,
mark options = {
    color = {rgb,1:red,0.00000000;green,0.00000000;blue,0.00000000}, draw opacity = 1.0,
    fill = {rgb,1:red,0.00000000;green,0.00000000;blue,0.00000000}, fill opacity = 1.0,
    line width = 1,
    rotate = 0,
    solid
},forget plot]coordinates {
(-0.51, 4.286263797015736e-17)
(-0.5099972740926235, 0.0013813499675772532)
(-0.5099890964129546, 0.002762678418399645)
(-0.5099754670883739, 0.004143963836048025)
(-0.50995638633118, 0.005525184704773042)
(-0.5099318544385866, 0.006906319509830802)
(-0.5099018717927173, 0.008287346737818457)
(-0.5098664388606003, 0.009668244877008375)
(-0.5098255561941608, 0.011048992417683697)
(-0.5097792244302122, 0.012429567852473842)
(-0.5097274442904465, 0.01380994967668843)
(-0.5096702165814229, 0.015190116388653507)
(-0.5096075421945553, 0.01657004649004537)
(-0.5095394221060984, 0.017949718486225907)
(-0.5094658573771326, 0.019329110886577877)
(-0.5093868491535471, 0.02070820220483873)
(-0.5093023986660227, 0.022086970959435754)
(-0.5092125072300118, 0.02346539567382116)
(-0.5091171762457185, 0.02484345487680566)
(-0.5090164071980771, 0.026221127102893406)
(-0.5089102016567276, 0.02759839089261681)
(-0.508798561275993, 0.02897522479286986)
(-0.5086814877948523, 0.030351607357242773)
(-0.5085589830369139, 0.03172751714635652)
(-0.508431048910387, 0.03310293272819584)
(-0.5082976874080521, 0.034477832678443544)
(-0.5081589006072299, 0.03585219558081471)
(-0.5080146906697486, 0.03722600002738934)
(-0.5078650598419107, 0.03859922461894628)
(-0.5077100104544578, 0.039971847965297)
(-0.5075495449225343, 0.04134384868561776)
(-0.5073836657456499, 0.042715205408783805)
(-0.5072123755076402, 0.04408589677370122)
(-0.5070356768766272, 0.045455901429640136)
(-0.5068535726049772, 0.04682519803656772)
(-0.5066660655292581, 0.04819376526547971)
(-0.5064731585701949, 0.04956158179873305)
(-0.5062748547326251, 0.05092862633037848)
(-0.5060711571054508, 0.05229487756649143)
(-0.5058620688615914, 0.05366031422550417)
(-0.5056475932579333, 0.05502491503853782)
(-0.5054277336352804, 0.056388658749732666)
(-0.5052024934183009, 0.057751524116579744)
(-0.5049718761154748, 0.05911348991025218)
(-0.5047358853190386, 0.06047453491593494)
(-0.5044945247049298, 0.06183463793315577)
(-0.5042477980327295, 0.06319377777611583)
(-0.5039957091456035, 0.06455193327401869)
(-0.5037382619702431, 0.06590908327140127)
(-0.5034754605168036, 0.06726520662846236)
(-0.5032073088788415, 0.06862028222139228)
(-0.5029338112332512, 0.06997428894270248)
(-0.5026549718402, 0.07132720570155332)
(-0.5023707950430615, 0.07267901142408308)
(-0.5020812852683475, 0.0740296850537367)
(-0.5017864470256401, 0.0753792055515928)
(-0.5014862849075208, 0.07672755189669195)
(-0.5011808035894986, 0.07807470308636445)
(-0.5008700078299382, 0.07942063813655664)
(-0.500553902469985, 0.08076533608215819)
(-0.5002324924334899, 0.0821087759773291)
(-0.499905782726933, 0.08345093689582513)
(-0.49957377843934514, 0.08479179793132413)
(-0.4992364847422288, 0.08613133819775218)
(-0.49889390688947755, 0.08746953682960788)
(-0.49854605021729415, 0.08880637298228856)
(-0.49819292014410765, 0.0901418258324139)
(-0.49783452217048874, 0.09147587457815073)
(-0.4974708618790641, 0.09280849843953755)
(-0.4971019449344296, 0.0941396766588073)
(-0.4967277770830619, 0.09546938850071106)
(-0.49634836415322925, 0.09679761325284161)
(-0.49596371205489986, 0.09812433022595513)
(-0.4955738267796512, 0.0994495187542938)
(-0.49517871440057526, 0.1007731581959083)
(-0.4947783810721851, 0.10209522793297839)
(-0.4943728330303182, 0.10341570737213451)
(-0.4939620765920396, 0.10473457594477899)
(-0.4935461181555437, 0.10605181310740555)
(-0.49312496420005436, 0.10736739834191979)
(-0.49269862128572395, 0.10868131115595914)
(-0.49226709605353136, 0.1099935310832113)
(-0.49183039522517835, 0.11130403768373338)
(-0.49138852560298485, 0.11261281054427077)
(-0.4909414940697835, 0.11391982927857405)
(-0.4904893075888115, 0.11522507352771776)
(-0.4900319732036029, 0.11652852296041646)
(-0.48956949803787897, 0.11783015727334185)
(-0.4891018892954364, 0.11912995619143951)
(-0.48862915426003595, 0.12042789946824385)
(-0.48815130029528875, 0.12172396688619382)
(-0.4876683348445411, 0.12301813825694836)
(-0.4871802654307593, 0.12431039342170001)
(-0.4866870996564122, 0.12560071225148928)
(-0.48618884520335215, 0.1268890746475186)
(-0.4856855098326964, 0.12817546054146467)
(-0.4851771013847057, 0.1294598498957914)
(-0.4846636277786617, 0.13074222270406233)
(-0.48414509701274433, 0.13202255899125168)
(-0.4836215171639072, 0.13330083881405572)
(-0.4830928963877512, 0.13457704226120384)
(-0.4825592429183979, 0.13585114945376783)
(-0.482020565068361, 0.13712314054547262)
(-0.4814768712284174, 0.13839299572300426)
(-0.4809281698674761, 0.1396606952063193)
(-0.4803744695324458, 0.140926219248953)
(-0.4798157788481028, 0.1421895481383262)
(-0.47925210651695616, 0.1434506621960529)
(-0.47868346131911177, 0.14470954177824696)
(-0.4781098521121362, 0.14596616727582745)
(-0.47753128783091847, 0.14722051911482448)
(-0.4769477774875307, 0.14847257775668438)
(-0.4763593301710882, 0.14972232369857344)
(-0.4757659550476073, 0.15096973747368186)
(-0.475167661359863, 0.1522147996515276)
(-0.47456445842724493, 0.15345749083825802)
(-0.4739563556456122, 0.1546977916769527)
(-0.4733433624871466, 0.15593568284792492)
(-0.4727254885002059, 0.157171145069022)
(-0.47210274330917434, 0.15840415909592662)
(-0.47147513661431306, 0.15963470572245564)
(-0.4708426781916092, 0.16086276578085956)
(-0.47020537789262296, 0.16208832014212174)
(-0.46956324564433505, 0.1633113497162553)
(-0.4689162914489914, 0.16453183545260103)
(-0.46826452538394736, 0.1657497583401245)
(-0.4676079576015111, 0.16696509940771143)
(-0.4669465983287854, 0.16817783972446346)
(-0.4662804578675077, 0.16938796039999357)
(-0.465609546593891, 0.17059544258471954)
(-0.46493387495846084, 0.17180026747015778)
(-0.4642534534858931, 0.17300241628921692)
(-0.4635682927748501, 0.17420187031648918)
(-0.4628784034978155, 0.17539861086854253)
(-0.4621837964009281, 0.17659261930421216)
(-0.4614844823038141, 0.17778387702488987)
(-0.4607804720994192, 0.17897236547481435)
(-0.4600717767538379, 0.1801580661413605)
(-0.4593584073061442, 0.18134096055532706)
(-0.45864037486821796, 0.182521030291225)
(-0.4579176906245731, 0.18369825696756395)
(-0.45719036583218287, 0.18487262224713877)
(-0.45645841182030455, 0.18604410783731554)
(-0.4557218399903029, 0.18721269549031574)
(-0.4549806618154727, 0.188378367003501)
(-0.45423488884085994, 0.18954110421965678)
(-0.453484532683082, 0.19070088902727472)
(-0.45272960503014703, 0.19185770336083488)
(-0.45197011764127104, 0.19301152920108763)
(-0.4512060823466959, 0.19416234857533363)
(-0.45043751104750407, 0.19531014355770404)
(-0.4496644157154335, 0.19645489626944024)
(-0.44888680839269157, 0.1975965888791714)
(-0.44810470119176715, 0.19873520360319283)
(-0.4473181062952418, 0.19987072270574321)
(-0.4465270359556004, 0.20100312849928)
(-0.4457315024950397, 0.202132403344756)
(-0.4449315183052771, 0.20325852965189323)
(-0.4441270958473573, 0.20438148987945715)
(-0.4433182476514579, 0.20550126653553055)
(-0.44250498631669477, 0.206617842177785)
(-0.4416873245109256, 0.20773119941375306)
(-0.4408652749705523, 0.2088413209010996)
(-0.44003885050032276, 0.209948189347891)
(-0.4392080639731317, 0.21105178751286507)
(-0.4383729283298199, 0.21215209820569994)
(-0.43753345657897247, 0.2132491042872809)
(-0.43668966179671653, 0.21434278866996812)
(-0.4358415571265173, 0.21543313431786276)
(-0.43498915577897346, 0.21652012424707187)
(-0.4341324710316117, 0.2176037415259733)
(-0.43327151622867877, 0.21868396927547962)
(-0.43240630478093545, 0.2197607906693004)
(-0.43153685016544585, 0.22083418893420506)
(-0.4306631659253686, 0.22190414735028355)
(-0.4297852656697456, 0.22297064925120696)
(-0.4289031630732897, 0.2240336780244875)
(-0.4280168718761723, 0.2250932171117367)
(-0.427126405883809, 0.22614925000892336)
(-0.4262317789666442, 0.2272017602666315)
(-0.4253330050599359, 0.22825073149031536)
(-0.42443009816353794, 0.22929614734055542)
(-0.42352307234168196, 0.23033799153331316)
(-0.42261194172275873, 0.23137624784018415)
(-0.42169672049909823, 0.2324109000886509)
(-0.42077742292674725, 0.2334419321623354)
(-0.419854063325249, 0.23446932800124937)
(-0.4189266560774194, 0.23549307160204466)
(-0.41799521562912256, 0.23651314701826306)
(-0.4170597564890465, 0.23752953836058388)
(-0.4161202932284769, 0.23854222979707185)
(-0.4151768404810696, 0.2395512055534239)
(-0.41422941294262383, 0.24055644991321443)
(-0.4132780253708519, 0.24155794721814072)
(-0.41232269258515064, 0.24255568186826615)
(-0.4113634294663697, 0.24354963832226345)
(-0.41040025095657995, 0.24453980109765727)
(-0.40943317205884067, 0.24552615477106457)
(-0.4084622078369665, 0.24650868397843523)
(-0.4074873734152916, 0.24748737341529164)
};
\addplot+ [color = {rgb,1:red,0.00000000;green,0.00000000;blue,0.00000000},
draw opacity = 1.0,
line width = 2.0,
solid,mark = none,
mark size = 2.0,
mark options = {
    color = {rgb,1:red,0.00000000;green,0.00000000;blue,0.00000000}, draw opacity = 1.0,
    fill = {rgb,1:red,0.50196078;green,0.50196078;blue,0.50196078}, fill opacity = 1.0,
    line width = 1,
    rotate = 0,
    solid
},fill = {rgb,1:red,0.50196078;green,0.50196078;blue,0.50196078}, fill opacity=1.0,forget plot]coordinates {
(0.0, 0.0)
(0.0, 0.24)
(-0.013333333333333334, 0.24)
(0.0, 0.26666666666666666)
(0.013333333333333334, 0.24)
(0.0, 0.24)
};
\addplot+ [color = {rgb,1:red,0.00000000;green,0.00000000;blue,0.00000000},
draw opacity = 1.0,
line width = 2.0,
solid,mark = none,
mark size = 2.0,
mark options = {
    color = {rgb,1:red,0.00000000;green,0.00000000;blue,0.00000000}, draw opacity = 1.0,
    fill = {rgb,1:red,0.50196078;green,0.50196078;blue,0.50196078}, fill opacity = 1.0,
    line width = 1,
    rotate = 0,
    solid
},fill = {rgb,1:red,0.50196078;green,0.50196078;blue,0.50196078}, fill opacity=1.0,forget plot]coordinates {
(0.0, 0.0)
(0.24, 0.0)
(0.24, 0.013333333333333334)
(0.26666666666666666, 0.0)
(0.24, -0.013333333333333334)
(0.24, 0.0)
};
\addplot+ [color = {rgb,1:red,0.00000000;green,0.00000000;blue,0.00000000},
draw opacity = 1.0,
line width = 1.0,
solid,mark = none,
mark size = 2.0,
mark options = {
    color = {rgb,1:red,0.00000000;green,0.00000000;blue,0.00000000}, draw opacity = 1.0,
    fill = {rgb,1:red,0.00000000;green,0.00000000;blue,0.00000000}, fill opacity = 1.0,
    line width = 1,
    rotate = 0,
    solid
},forget plot]coordinates {
(0.0, 0.08888888888888889)
(0.08888888888888889, 0.08888888888888889)
};
\addplot+ [color = {rgb,1:red,0.00000000;green,0.00000000;blue,0.00000000},
draw opacity = 1.0,
line width = 1.0,
solid,mark = none,
mark size = 2.0,
mark options = {
    color = {rgb,1:red,0.00000000;green,0.00000000;blue,0.00000000}, draw opacity = 1.0,
    fill = {rgb,1:red,0.00000000;green,0.00000000;blue,0.00000000}, fill opacity = 1.0,
    line width = 1,
    rotate = 0,
    solid
},forget plot]coordinates {
(0.08888888888888889, 0.0)
(0.08888888888888889, 0.08888888888888889)
};
\addplot+[draw=none, color = {rgb,1:red,0.42314674;green,0.62249549;blue,0.19877060},
draw opacity = 1.0,
line width = 0,
solid,mark = *,
mark size = 0.5333333333333333,
mark options = {
    color = {rgb,1:red,0.00000000;green,0.00000000;blue,0.00000000}, draw opacity = 1.0,
    fill = {rgb,1:red,0.00000000;green,0.00000000;blue,0.00000000}, fill opacity = 0.0,
    line width = 1,
    rotate = 0,
    solid
},forget plot] coordinates {
(0.044444444444444446, 0.044444444444444446)
};
\node at (axis cs:1.6882898355963438, 0.0976957051531751) [,
color={rgb,1:red,0.00000000;green,0.00000000;blue,0.00000000}, draw opacity=1.0,
rotate=0.0,
font={\fontsize{14 pt}{18.2 pt}\selectfont}
] {\textcolor{gray}{Ensembled averaged}$$};
\node at (axis cs:1.6882898355963438, -0.0976957051531751) [,
color={rgb,1:red,0.00000000;green,0.00000000;blue,0.00000000}, draw opacity=1.0,
rotate=0.0,
font={\fontsize{14 pt}{18.2 pt}\selectfont}
] {\textcolor{gray}{particulate material}$$};
\node at (axis cs:0.40382918157083725, -0.958047144249125) [,
color={rgb,1:red,0.00000000;green,0.00000000;blue,0.00000000}, draw opacity=1.0,
rotate=0.0,
font={\fontsize{14 pt}{18.2 pt}\selectfont}
] {$ \mathrm{Re} \, \mathbf s_2 $};
\node at (axis cs:0.384533850361305, -0.4171246727399496) [,
color={rgb,1:red,0.00000000;green,0.00000000;blue,0.00000000}, draw opacity=1.0,
rotate=0.0,
font={\fontsize{14 pt}{18.2 pt}\selectfont}
] {$ \mathrm{Re} \, \mathbf s_0 $};
\node at (axis cs:0.5719580117363765, 0.9485980639478271) [,
color={rgb,1:red,0.00000000;green,0.00000000;blue,0.00000000}, draw opacity=1.0,
rotate=0.0,
font={\fontsize{14 pt}{18.2 pt}\selectfont}
] {$ \mathrm{Re} \, \mathbf s_1 $};
\node at (axis cs:0.44505740354764783, 1.414792034862466) [,
color={rgb,1:red,0.00000000;green,0.00000000;blue,0.00000000}, draw opacity=1.0,
rotate=0.0,
font={\fontsize{14 pt}{18.2 pt}\selectfont}
] {$ \mathrm{Re} \, \mathbf s_3 $};
\node at (axis cs:-0.38355339059327376, -0.4228867239266071) [,
color={rgb,1:red,0.00000000;green,0.00000000;blue,0.00000000}, draw opacity=1.0,
rotate=0.0,
font={\fontsize{14 pt}{18.2 pt}\selectfont}
] {$ \mathbf k $};
\node at (axis cs:-0.684466666666666, -0.23266666666666666) [,
color={rgb,1:red,0.00000000;green,0.00000000;blue,0.00000000}, draw opacity=1.0,
rotate=0.0,
font={\fontsize{14 pt}{18.2 pt}\selectfont}
] {$\theta_\mathrm{in}$};
\node at (axis cs:-0.684401228643623, 0.2314276634023983) [,
color={rgb,1:red,0.00000000;green,0.00000000;blue,0.00000000}, draw opacity=1.0,
rotate=0.0,
font={\fontsize{14 pt}{18.2 pt}\selectfont}
] {$\theta_\mathrm{in}$};
\node at (axis cs:-0.13866666666666666, 0.26666666666666666) [,
color={rgb,1:red,0.00000000;green,0.00000000;blue,0.00000000}, draw opacity=1.0,
rotate=0.0,
font={\fontsize{14 pt}{18.2 pt}\selectfont}
] {$y$};
\node at (axis cs:0.26666666666666666, -0.13866666666666666) [,
color={rgb,1:red,0.00000000;green,0.00000000;blue,0.00000000}, draw opacity=1.0,
rotate=0.0,
font={\fontsize{14 pt}{18.2 pt}\selectfont}
] {$x$};
\end{axis}

\end{tikzpicture}

%% file: images/wavenumbers-example.tex
\begin{tikzpicture}[]
\begin{axis}[height = {53.34mm}, ylabel = {Im $S$}, xmin = {-81.11240180766393}, xmax = {76.77140483745376}, ymax = {6.924846918374815}, xlabel = {Re $S$}, unbounded coords=jump,scaled x ticks = false,xlabel style = {font = {\fontsize{11 pt}{14.3 pt}\selectfont}, color = {rgb,1:red,0.00000000;green,0.00000000;blue,0.00000000}, draw opacity = 1.0, rotate = 0.0},xmajorgrids = true,xtick = {-60.0,-30.0,0.0,30.0,60.0},xticklabels = {$-60$,$-30$,$0$,$30$,$60$},xtick align = inside,xticklabel style = {font = {\fontsize{8 pt}{10.4 pt}\selectfont}, color = {rgb,1:red,0.00000000;green,0.00000000;blue,0.00000000}, draw opacity = 1.0, rotate = 0.0},x grid style = {color = {rgb,1:red,0.00000000;green,0.00000000;blue,0.00000000},
draw opacity = 0.1,
line width = 0.5,
solid},axis lines* = left,x axis line style = {color = {rgb,1:red,0.00000000;green,0.00000000;blue,0.00000000},
draw opacity = 1.0,
line width = 1,
solid},scaled y ticks = false,ylabel style = {font = {\fontsize{11 pt}{14.3 pt}\selectfont}, color = {rgb,1:red,0.00000000;green,0.00000000;blue,0.00000000}, draw opacity = 1.0, rotate = 0.0},ymajorgrids = true,ytick = {0.0,1.0,2.0,3.0,4.0,5.0,6.0},yticklabels = {$0$,$1$,$2$,$3$,$4$,$5$,$6$},ytick align = inside,yticklabel style = {font = {\fontsize{8 pt}{10.4 pt}\selectfont}, color = {rgb,1:red,0.00000000;green,0.00000000;blue,0.00000000}, draw opacity = 1.0, rotate = 0.0},y grid style = {color = {rgb,1:red,0.00000000;green,0.00000000;blue,0.00000000},
draw opacity = 0.1,
line width = 0.5,
solid},axis lines* = left,y axis line style = {color = {rgb,1:red,0.00000000;green,0.00000000;blue,0.00000000},
draw opacity = 1.0,
line width = 1,
solid},    xshift = 0.0mm,
    yshift = 0.0mm,
    axis background/.style={fill={rgb,1:red,1.00000000;green,1.00000000;blue,1.00000000}}
,legend style = {color = {rgb,1:red,0.00000000;green,0.00000000;blue,0.00000000},
draw opacity = 1.0,
line width = 1,
solid,fill = {rgb,1:red,1.00000000;green,1.00000000;blue,1.00000000},font = {\fontsize{8 pt}{10.4 pt}\selectfont}},colorbar style={title=}, ymin = {0.0}, width = {65.024mm}]\addplot+[draw=none, color = {rgb,1:red,0.00000000;green,0.60560316;blue,0.97868012},
draw opacity = 1.0,
line width = 0,
solid,mark = *,
mark size = 1.25,
mark options = {
    color = {rgb,1:red,0.00000000;green,0.00000000;blue,0.00000000}, draw opacity = 1.0,
    fill = {rgb,1:red,0.00000000;green,0.00000000;blue,1.00000000}, fill opacity = 1.0,
    line width = 0.3,
    rotate = 0,
    solid
},forget plot] coordinates {
(2.0370403700602835, 1.246171577286332)
(0.8252138771844854, 3.34261471028739)
(8.96835251479123, 3.685046716670963)
(15.445799942323548, 4.488431976070869)
(8.441447349566108, 4.875032390776338)
(21.826334148501555, 5.004578707794722)
(28.169471568543326, 5.387103862038009)
(14.983765560936172, 5.632459514960552)
(34.493780390123526, 5.691453980222071)
(6.407666212804686, 5.801314574879608)
(40.80714248894949, 5.94430791775453)
(21.391316003747644, 6.133073393657205)
(47.113526153846436, 6.160625131486249)
(53.41516431228585, 6.349654664314246)
(27.74940487956722, 6.507936035692962)
(59.713416738327304, 6.517520591247036)
(13.091237556548549, 6.603113096600146)
(66.00916212255402, 6.668490000223288)
(72.30299521542213, 6.805656007079733)
(34.08325709197962, 6.807761307223201)
};
\addplot+[draw=none, color = {rgb,1:red,0.88887350;green,0.43564919;blue,0.27812294},
draw opacity = 1.0,
line width = 0,
solid,mark = *,
mark size = 1.25,
mark options = {
    color = {rgb,1:red,0.00000000;green,0.00000000;blue,0.00000000}, draw opacity = 1.0,
    fill = {rgb,1:red,1.00000000;green,0.00000000;blue,0.00000000}, fill opacity = 1.0,
    line width = 0.3,
    rotate = 0,
    solid
},forget plot] coordinates {
(-0.08258507892237334, 1.1602383294359326)
(-6.99832635109368, 3.6280345067739006)
(-13.48708080386335, 4.455287612032784)
(-7.1656662474885335, 4.774457263488311)
(-19.87105589638136, 4.981125702532772)
(-1.6176179069248537, 5.168219272695003)
(-26.215739833009827, 5.368939990493103)
(-13.73597147744943, 5.568133607881797)
(-32.540888048216374, 5.676628586154263)
(-38.85476081701204, 5.931783264764539)
(-20.154120840710227, 6.083977904303744)
(-45.16148011239225, 6.1497827968163055)
(-51.46335156819863, 6.340096206882568)
(-26.51799291856338, 6.468040251353798)
(-57.76177321536864, 6.508974196132058)
(-64.05764552794963, 6.660761931844391)
(-32.85532894065895, 6.774217327386513)
(-70.351576441911, 6.798603305951178)
(-15.420319816331771, 6.846248365660278)
(-76.6439921856323, 6.924846918374815)
};
\addplot+ [color = {rgb,1:red,0.50196078;green,0.50196078;blue,0.50196078},
draw opacity = 1.0,
line width = 2,
dashed,mark = none,
mark size = 2.0,
mark options = {
    color = {rgb,1:red,0.00000000;green,0.00000000;blue,0.00000000}, draw opacity = 1.0,
    fill = {rgb,1:red,0.50196078;green,0.50196078;blue,0.50196078}, fill opacity = 1.0,
    line width = 1,
    rotate = 0,
    solid
},forget plot]coordinates {
(9.455729441499583, 1.2032049533611322)
(9.451503664010737, 1.219481681428435)
(9.438830543891097, 1.2357421844992196)
(9.41772271398241, 1.2519702537503974)
(9.388201215028575, 1.2681497126896164)
(9.350295474701776, 1.2842644332803408)
(9.304043278268324, 1.3002983520186282)
(9.249490730923506, 1.3162354859455805)
(9.186692211832904, 1.3320599485795035)
(9.115710319926068, 1.3477559657518996)
(9.03661581149654, 1.363307891331499)
(8.949487529670442, 1.3787002228206668)
(8.854412325813936, 1.3939176168086271)
(8.751484972957885, 1.4089449042661102)
(8.640808071326047, 1.4237671056661703)
(8.522491946060931, 1.4383694459161038)
(8.396654537249297, 1.4527373690855827)
(8.26342128235692, 1.4668565529163233)
(8.122924991189786, 1.4807129230988236)
(7.975305713506393, 1.494292667301943)
(7.820710599413111, 1.5075822489413315)
(7.65929375268175, 1.5205684206729941)
(7.491216077135566, 1.5332382375985296)
(7.316645116256839, 1.5455790701688883)
(7.135754886175883, 1.5575786167737797)
(6.94872570220798, 1.569224916004186)
(6.755743999111155, 1.580506358575755)
(6.5570021452439535, 1.5914116989011862)
(6.352698250808479, 1.6019300663000773)
(6.143035970369835, 1.612050975835056)
(5.928224299848827, 1.6217643387633913)
(5.708477368190277, 1.631060472593674)
(5.4840142239146425, 1.6399301107375321)
(5.2550586167656865, 1.6483644117467706)
(5.021838774671859, 1.6563549681267171)
(4.784587176243746, 1.6638938147169964)
(4.5435403190343235, 1.6709734366313747)
(4.298938483793061, 1.6775867767487629)
(4.051025494948846, 1.6837272427479073)
(3.8000484775604795, 1.689388713678759)
(3.546257610977059, 1.6945655460639708)
(3.28990587945375, 1.699252579524437)
(3.0312488199715792, 1.7034451419232726)
(2.770544267512627, 1.707139054023099)
(2.5080520980444945, 1.7103306336520023)
(2.244033969470295, 1.7130166993739986)
(1.9787530608023687, 1.7151945736603604)
(1.712473809819711, 1.7168620855586358)
(1.4454616494706483, 1.7180175728567)
(1.1779827432835077, 1.7186598837396847)
(0.9103037200490114, 1.7187883779381337)
(0.6426914080389134, 1.7184029273662351)
(0.37541256902576325, 1.7175039162495018)
(0.1087336323689746, 1.716092240741766)
(-0.15707957056774857, 1.714169308031875)
(-0.4217620714029071, 1.711737034940973)
(-0.6850500288634482, 1.708797846011773)
(-0.9466809917874761, 1.7053546710917205)
(-1.2063941607412925, 1.7014109424124566)
(-1.4639306479899286, 1.6969705911684956)
(-1.719033735562083, 1.6920380435985245)
(-1.9714491311521727, 1.6866182165732329)
(-2.2209252216044275, 1.6807165126940677)
(-2.467213323726366, 1.674338814907804)
(-2.710067932181608, 1.6674914806422954)
(-2.9492469642149226, 1.6601813354692505)
(-3.1845120009656007, 1.6524156663003569)
(-3.415628525128546, 1.6442022141235269)
(-3.6423661547262327, 1.6355491662865136)
(-3.864498872758462, 1.626465148335585)
(-4.081805252501037, 1.61695921541739)
(-4.294068678228727, 1.6070408432525936)
(-4.501077561142534, 1.5967199186902705)
(-4.702625550286017, 1.5860067298524807)
(-4.898511738240394, 1.5749119558788445)
(-5.088540861393453, 1.5634466562813438)
(-5.272523494582546, 1.551622259919961)
(-5.450276239917721, 1.5394505536101417)
(-5.621621909596724, 1.5269436703734418)
(-5.7863897025296485, 1.5141140773430668)
(-5.94441537459719, 1.5009745633363631)
(-6.095541402372741, 1.4875382261066463)
(-6.2396171401451825, 1.4738184592870756)
(-6.376498970085791, 1.459828939039589)
(-6.506050445409625, 1.445583610422209)
(-6.6281424263886395, 1.431096673488304)
(-6.742653209080982, 1.4163825691316663)
(-6.849468646648122, 1.4014559646915157)
(-6.94848226313891, 1.386331739331777)
(-7.039595359627115, 1.3710249692092045)
(-7.122717112596658, 1.3555509124451437)
(-7.197764664476478, 1.339924993915901)
(-7.264663206234753, 1.3241627898768935)
(-7.323346051950174, 1.308280012435898)
(-7.373754705285914, 1.2922924938908809)
(-7.41583891780006, 1.2762161709480184)
(-7.449556739034347, 1.2600670688356432)
(-7.474874558331281, 1.2438612853299456)
(-7.491767138337987, 1.2276149747083616)
(-7.500217640163339, 1.211344331646638)
(-7.500217640163339, 1.1950655750756267)
(-7.491767138337987, 1.178794932013903)
(-7.474874558331281, 1.162548621392319)
(-7.449556739034347, 1.1463428378866212)
(-7.415838917800062, 1.130193735774246)
(-7.373754705285914, 1.1141174128313838)
(-7.323346051950174, 1.0981298942863666)
(-7.264663206234753, 1.0822471168453711)
(-7.1977646644764794, 1.0664849128063634)
(-7.12271711259666, 1.0508589942771207)
(-7.039595359627115, 1.03538493751306)
(-6.948482263138912, 1.0200781673904877)
(-6.849468646648123, 1.004953942030749)
(-6.742653209080982, 0.9900273375905984)
(-6.628142426388641, 0.9753132332339605)
(-6.506050445409626, 0.9608262963000553)
(-6.376498970085792, 0.9465809676826753)
(-6.239617140145183, 0.9325914474351891)
(-6.095541402372742, 0.9188716806156182)
(-5.944415374597191, 0.9054353433859013)
(-5.78638970252965, 0.8922958293791976)
(-5.621621909596725, 0.8794662363488228)
(-5.450276239917723, 0.8669593531121228)
(-5.272523494582548, 0.8547876468023035)
(-5.088540861393454, 0.8429632504409207)
(-4.898511738240396, 0.83149795084342)
(-4.702625550286018, 0.8204031768697837)
(-4.501077561142539, 0.8096899880319941)
(-4.2940686782287285, 0.799369063469671)
(-4.081805252501042, 0.7894506913048747)
(-3.864498872758464, 0.7799447583866795)
(-3.642366154726232, 0.7708607404357506)
(-3.4156285251285503, 0.7622076925987378)
(-3.1845120009656025, 0.7539942404219075)
(-2.949246964214928, 0.7462285712530141)
(-2.7100679321816097, 0.7389184260799693)
(-2.467213323726366, 0.7320710918144603)
(-2.22092522160443, 0.7256933940281967)
(-1.9714491311521731, 0.7197916901490314)
(-1.7190337355620888, 0.7143718631237399)
(-1.4639306479899308, 0.7094393155537689)
(-1.2063941607412907, 0.7049989643098078)
(-0.9466809917874799, 0.701055235630544)
(-0.6850500288634485, 0.6976120607104914)
(-0.4217620714029131, 0.6946728717812916)
(-0.15707957056775035, 0.6922405986903895)
(0.10873363236897626, 0.6903176659804983)
(0.37541256902575937, 0.6889059904727627)
(0.6426914080389132, 0.6880069793560293)
(0.9103037200490074, 0.6876215287841307)
(1.1779827432835057, 0.6877500229825797)
(1.4454616494706498, 0.6883923338655645)
(1.712473809819707, 0.6895478211636286)
(1.9787530608023687, 0.691215333061904)
(2.244033969470291, 0.6933932073482658)
(2.5080520980444927, 0.696079273070262)
(2.770544267512628, 0.6992708526991652)
(3.0312488199715766, 0.702964764798992)
(3.289905879453749, 0.7071573271978273)
(3.5462576109770554, 0.7118443606582934)
(3.8000484775604795, 0.7170211930435053)
(4.05102549494884, 0.722682663974357)
(4.2989384837930595, 0.7288231299735015)
(4.5435403190343235, 0.7354364700908897)
(4.784587176243742, 0.7425160920052679)
(5.021838774671859, 0.7500549385955473)
(5.255058616765682, 0.7580454949754936)
(5.484014223914641, 0.7664797959847323)
(5.708477368190278, 0.7753494341285905)
(5.928224299848824, 0.7846455679588729)
(6.143035970369835, 0.7943589308872083)
(6.3526982508084755, 0.8044798404221868)
(6.557002145243952, 0.8149982078210782)
(6.755743999111156, 0.8259035481465095)
(6.9487257022079785, 0.8371849907180782)
(7.135754886175882, 0.8488312899484847)
(7.316645116256836, 0.8608308365533759)
(7.491216077135563, 0.8731716691237347)
(7.659293752681751, 0.8858414860492705)
(7.820710599413109, 0.8988276577809328)
(7.975305713506393, 0.9121172394203214)
(8.122924991189784, 0.9256969836234403)
(8.26342128235692, 0.9395533538059411)
(8.396654537249299, 0.9536725376366818)
(8.522491946060928, 0.9680404608061605)
(8.640808071326047, 0.9826428010560941)
(8.751484972957883, 0.9974650024561539)
(8.854412325813934, 1.0124922899136373)
(8.94948752967044, 1.0277096839015973)
(9.03661581149654, 1.043102015390765)
(9.115710319926068, 1.0586539409703648)
(9.186692211832904, 1.0743499581427605)
(9.249490730923506, 1.090174420776684)
(9.304043278268324, 1.106111554703636)
(9.350295474701774, 1.1221454734419234)
(9.388201215028577, 1.138260194032648)
(9.41772271398241, 1.1544396529718668)
(9.438830543891097, 1.1706677222230448)
(9.451503664010735, 1.1869282252938291)
(9.455729441499583, 1.203204953361132)
};
\node at (axis cs:-41.415281334084185, 2.2344039275659244) [,
color={rgb,1:red,0.00000000;green,0.00000000;blue,0.00000000}, draw opacity=1.0,
rotate=0.0,
font={\fontsize{14 pt}{18.2 pt}\selectfont}
] {\textcolor{red}{backward}$$};
\node at (axis cs:43.36973662522209, 2.2344039275659244) [,
color={rgb,1:red,0.00000000;green,0.00000000;blue,0.00000000}, draw opacity=1.0,
rotate=0.0,
font={\fontsize{14 pt}{18.2 pt}\selectfont}
] {\textcolor{blue}{forward}$$};
\end{axis}
\begin{axis}[height = {53.34mm}, ylabel = {}, xmin = {-6.427446662976433}, xmax = {8.833856569698698}, ymax = {1.5125646456225699}, xlabel = {Re $S$}, unbounded coords=jump,scaled x ticks = false,xlabel style = {font = {\fontsize{11 pt}{14.3 pt}\selectfont}, color = {rgb,1:red,0.00000000;green,0.00000000;blue,0.00000000}, draw opacity = 1.0, rotate = 0.0},xmajorgrids = true,xtick = {-6.0,-3.0,0.0,3.0,6.0},xticklabels = {$-6$,$-3$,$0$,$3$,$6$},xtick align = inside,xticklabel style = {font = {\fontsize{8 pt}{10.4 pt}\selectfont}, color = {rgb,1:red,0.00000000;green,0.00000000;blue,0.00000000}, draw opacity = 1.0, rotate = 0.0},x grid style = {color = {rgb,1:red,0.00000000;green,0.00000000;blue,0.00000000},
draw opacity = 0.1,
line width = 0.5,
solid},axis lines* = left,x axis line style = {color = {rgb,1:red,0.00000000;green,0.00000000;blue,0.00000000},
draw opacity = 1.0,
line width = 1,
solid},scaled y ticks = false,ylabel style = {font = {\fontsize{11 pt}{14.3 pt}\selectfont}, color = {rgb,1:red,0.00000000;green,0.00000000;blue,0.00000000}, draw opacity = 1.0, rotate = 0.0},ymajorgrids = true,ytick = {0.0,0.5,1.0,1.5},yticklabels = {$0.0$,$0.5$,$1.0$,$1.5$},ytick align = inside,yticklabel style = {font = {\fontsize{8 pt}{10.4 pt}\selectfont}, color = {rgb,1:red,0.00000000;green,0.00000000;blue,0.00000000}, draw opacity = 1.0, rotate = 0.0},y grid style = {color = {rgb,1:red,0.00000000;green,0.00000000;blue,0.00000000},
draw opacity = 0.1,
line width = 0.5,
solid},axis lines* = left,y axis line style = {color = {rgb,1:red,0.00000000;green,0.00000000;blue,0.00000000},
draw opacity = 1.0,
line width = 1,
solid},    xshift = 65.024mm,
    yshift = 0.0mm,
    axis background/.style={fill={rgb,1:red,1.00000000;green,1.00000000;blue,1.00000000}}
,legend style = {color = {rgb,1:red,0.00000000;green,0.00000000;blue,0.00000000},
draw opacity = 1.0,
line width = 1,
solid,fill = {rgb,1:red,1.00000000;green,1.00000000;blue,1.00000000},font = {\fontsize{8 pt}{10.4 pt}\selectfont}},colorbar style={title=}, ymin = {0.0}, width = {65.65899999999999mm}]\addplot+[draw=none, color = {rgb,1:red,0.00000000;green,0.60560316;blue,0.97868012},
draw opacity = 1.0,
line width = 0,
solid,mark = *,
mark size = 1.7,
mark options = {
    color = {rgb,1:red,0.00000000;green,0.00000000;blue,0.00000000}, draw opacity = 1.0,
    fill = {rgb,1:red,0.00000000;green,0.00000000;blue,1.00000000}, fill opacity = 1.0,
    line width = 0.3,
    rotate = 0,
    solid
},forget plot] coordinates {
(2.0370403700602835, 1.246171577286332)
};
\addplot+[draw=none, color = {rgb,1:red,0.88887350;green,0.43564919;blue,0.27812294},
draw opacity = 1.0,
line width = 0,
solid,mark = *,
mark size = 1.7,
mark options = {
    color = {rgb,1:red,0.00000000;green,0.00000000;blue,0.00000000}, draw opacity = 1.0,
    fill = {rgb,1:red,1.00000000;green,0.00000000;blue,0.00000000}, fill opacity = 1.0,
    line width = 0.3,
    rotate = 0,
    solid
},forget plot] coordinates {
(-0.08258507892237334, 1.1602383294359326)
};
\addplot+ [color = {rgb,1:red,0.50196078;green,0.50196078;blue,0.50196078},
draw opacity = 1.0,
line width = 2,
dashed,mark = none,
mark size = 2.0,
mark options = {
    color = {rgb,1:red,0.00000000;green,0.00000000;blue,0.00000000}, draw opacity = 1.0,
    fill = {rgb,1:red,0.50196078;green,0.50196078;blue,0.50196078}, fill opacity = 1.0,
    line width = 1,
    rotate = 0,
    solid
},forget plot]coordinates {
(5.216478543534269, 1.2032049533611322)
(5.214365654789845, 1.2113433173947836)
(5.208029094730025, 1.2194735689301759)
(5.197475179775682, 1.227587603555765)
(5.182714430298764, 1.2356773330253743)
(5.163761560135365, 1.2437346933207365)
(5.140635461918639, 1.2517516526898802)
(5.11335918824623, 1.2597202196533563)
(5.081959928700929, 1.2676324509703178)
(5.046468982747511, 1.275480459556516)
(5.006921728532747, 1.2832564223463157)
(4.963357587619699, 1.2909525880908994)
(4.915819985691445, 1.2985612850848796)
(4.86435630926342, 1.3060749288136213)
(4.8090178584475005, 1.3134860295136512)
(4.749859795814943, 1.320787199638618)
(4.6869410914091265, 1.3279711612233573)
(4.6203244639629375, 1.3350307531387278)
(4.55007631837937, 1.341958938229978)
(4.476266679537674, 1.3487488103315375)
(4.3989691224910334, 1.3553936011512318)
(4.318260699125353, 1.3618866870170632)
(4.23422186135226, 1.368221595479831)
(4.146936380912897, 1.3743920117650104)
(4.0564912658724195, 1.3803917850674559)
(3.962976673888468, 1.386214934682659)
(3.8664858223400556, 1.3918556559684436)
(3.767114895406454, 1.3973083261311592)
(3.664962948188717, 1.4025675098306047)
(3.5601318079693955, 1.407627964598094)
(3.4527259727088913, 1.4124846460622618)
(3.342852506879616, 1.417132712977403)
(3.2306209347417987, 1.4215675320493322)
(3.1161431311673207, 1.4257846825539513)
(2.999533210120407, 1.4297799607439248)
(2.8809074109063504, 1.4335493840390643)
(2.760383982301639, 1.4370891949962534)
(2.638083064681008, 1.4403958650549475)
(2.5141265702589006, 1.4434660980545198)
(2.388638061564717, 1.4462968335199458)
(2.261742628273007, 1.4488852497125515)
(2.1335667625113524, 1.4512287664427848)
(2.004238232770267, 1.4533250476422024)
(1.873885956540791, 1.4551720036921156)
(1.7426398718067249, 1.4567677935065673)
(1.610630807519625, 1.4581108263675655)
(1.4779903531856617, 1.4591997635107463)
(1.3448507276943331, 1.4600335194598841)
(1.2113446475198018, 1.460611263108916)
(1.0776051944262313, 1.4609324185504085)
(0.9437656828089832, 1.460996665649633)
(0.8099595268039342, 1.4608039403636837)
(0.6763201072973591, 1.460354434805317)
(0.5429806389689649, 1.4596485970514492)
(0.41007403750060323, 1.4586871306965037)
(0.27773278708302396, 1.4574709941510526)
(0.1460888083527534, 1.4560013996864525)
(0.015273326890739458, 1.4542798122264262)
(-0.11458325758616872, 1.4523079478867944)
(-0.24335150121048676, 1.4500877722648138)
(-0.370903044996564, 1.4476214984798284)
(-0.49711074279160883, 1.4449115849671825)
(-0.6218487880177361, 1.4419607330276)
(-0.7449928390787055, 1.438771884134468)
(-0.8664201433063263, 1.4353482170017138)
(-0.9860096593229839, 1.4316931444151915)
(-1.1036421776983227, 1.4278103098307446)
(-1.2192004397797953, 1.4237035837423295)
(-1.3325692545786387, 1.419377059823823)
(-1.4436356135947535, 1.4148350508483585)
(-1.552288803466041, 1.410082084389261)
(-1.6584205163298857, 1.4051228983068629)
(-1.7619249577867893, 1.3999624360257013)
(-1.862698952358531, 1.3946058416068063)
(-1.9606420463357195, 1.3890584546199882)
(-2.0556566079122485, 1.383325804821238)
(-2.1476479245067956, 1.3774136066405465)
(-2.236524297174383, 1.371327753485637)
(-2.3221971320138843, 1.365074311867287)
(-2.4045810284803464, 1.3586595153520995)
(-2.483593864514117, 1.3520897583487477)
(-2.559156878401893, 1.3453715897338894)
(-2.631194747288114, 1.3385117063241039)
(-2.6996356622584177, 1.3315169462003607)
(-2.764411399920335, 1.3243942818916707)
(-2.8254573904098423, 1.317150813424718)
(-2.8827127817560134, 1.3097937612463992)
(-2.936120500539583, 1.3023304590263238)
(-2.985627308784977, 1.2947683463464545)
(-3.03118385702908, 1.2871149612851684)
(-3.0727447335138516, 1.279377932903138)
(-3.1102685094537614, 1.2715649736385166)
(-3.1437177803328993, 1.2636838716190129)
(-3.1730592031906095, 1.255742482898515)
(-3.1982635298584796, 1.2477487236260065)
(-3.2193056361155525, 1.2397105621545754)
(-3.236164546732696, 1.2316360110983877)
(-3.248823456381163, 1.2235331193455388)
(-3.257269746384516, 1.2154099640347469)
(-3.2614949972971923, 1.207274642503885)
(-3.2614949972971923, 1.1991352642183795)
(-3.257269746384516, 1.1909999426875175)
(-3.248823456381163, 1.1828767873767256)
(-3.236164546732696, 1.1747738956238767)
(-3.2193056361155534, 1.1666993445676892)
(-3.1982635298584796, 1.1586611830962579)
(-3.1730592031906095, 1.1506674238237493)
(-3.1437177803328993, 1.1427260351032515)
(-3.1102685094537623, 1.1348449330837478)
(-3.0727447335138525, 1.1270319738191266)
(-3.03118385702908, 1.1192949454370962)
(-2.985627308784978, 1.1116415603758099)
(-2.936120500539584, 1.1040794476959406)
(-2.8827127817560134, 1.0966161454758652)
(-2.8254573904098432, 1.0892590932975463)
(-2.764411399920335, 1.0820156248305939)
(-2.6996356622584186, 1.0748929605219038)
(-2.631194747288114, 1.0678982003981605)
(-2.559156878401893, 1.0610383169883753)
(-2.483593864514118, 1.0543201483735167)
(-2.4045810284803473, 1.047750391370165)
(-2.3221971320138852, 1.0413355948549774)
(-2.236524297174384, 1.0350821532366274)
(-2.1476479245067965, 1.028996300081718)
(-2.0556566079122494, 1.0230841019010264)
(-1.9606420463357204, 1.0173514521022762)
(-1.8626989523585313, 1.011804065115458)
(-1.761924957786792, 1.0064474706965632)
(-1.6584205163298866, 1.0012870084154015)
(-1.5522888034660436, 0.9963278223330034)
(-1.4436356135947543, 0.9915748558739059)
(-1.3325692545786383, 0.9870328468984414)
(-1.2192004397797975, 0.982706322979935)
(-1.1036421776983236, 0.9785995968915199)
(-0.9860096593229866, 0.9747167623070732)
(-0.8664201433063272, 0.9710616897205507)
(-0.7449928390787055, 0.9676380225877963)
(-0.6218487880177377, 0.9644491736946644)
(-0.49711074279160905, 0.9614983217550819)
(-0.3709030449965669, 0.958788408242436)
(-0.24335150121048787, 0.9563221344574506)
(-0.11458325758616783, 0.95410195883547)
(0.01527332689073757, 0.9521300944958381)
(0.14608880835275329, 0.9504085070358118)
(0.27773278708302096, 0.9489389125712119)
(0.41007403750060234, 0.9477227760257608)
(0.5429806389689656, 0.9467613096708152)
(0.6763201072973573, 0.9460554719169474)
(0.8099595268039341, 0.9456059663585807)
(0.9437656828089812, 0.9454132410726315)
(1.0776051944262304, 0.945477488171856)
(1.2113446475198024, 0.9457986436133483)
(1.3448507276943311, 0.9463763872623804)
(1.4779903531856617, 0.9472101432115181)
(1.6106308075196232, 0.948299080354699)
(1.742639871806724, 0.9496421132156971)
(1.8738859565407917, 0.9512379030301488)
(2.004238232770266, 0.953084859080062)
(2.133566762511352, 0.9551811402794798)
(2.261742628273005, 0.9575246570097128)
(2.388638061564717, 0.9601130732023188)
(2.514126570258898, 0.9629438086677446)
(2.638083064681007, 0.9660140416673169)
(2.760383982301639, 0.969320711726011)
(2.8809074109063486, 0.9728605226832001)
(2.999533210120407, 0.9766299459783397)
(3.116143131167319, 0.9806252241683129)
(3.230620934741798, 0.9848423746729322)
(3.342852506879616, 0.9892771937448613)
(3.4527259727088895, 0.9939252606600025)
(3.5601318079693955, 0.9987819421241703)
(3.664962948188715, 1.0038423968916594)
(3.7671148954064533, 1.0091015805911052)
(3.8664858223400556, 1.0145542507538208)
(3.962976673888467, 1.020194972039605)
(4.056491265872419, 1.0260181216548085)
(4.146936380912895, 1.032017894957254)
(4.2342218613522595, 1.0381883112424335)
(4.318260699125353, 1.0445232197052015)
(4.398969122491033, 1.0510163055710324)
(4.476266679537674, 1.057661096390727)
(4.550076318379369, 1.0644509684922863)
(4.620324463962937, 1.0713791535835366)
(4.6869410914091265, 1.078438745498907)
(4.749859795814942, 1.0856227070836464)
(4.8090178584475005, 1.0929238772086132)
(4.864356309263419, 1.100334977908643)
(4.915819985691445, 1.1078486216373846)
(4.963357587619698, 1.1154573186313648)
(5.006921728532747, 1.1231534843759488)
(5.046468982747511, 1.1309294471657485)
(5.081959928700929, 1.1387774557519463)
(5.11335918824623, 1.146689687068908)
(5.140635461918639, 1.154658254032384)
(5.163761560135364, 1.162675213401528)
(5.182714430298765, 1.17073257369689)
(5.197475179775682, 1.1788223031664995)
(5.208029094730025, 1.1869363377920885)
(5.2143656547898445, 1.1950665893274806)
(5.216478543534269, 1.2032049533611322)
};
\node at (axis cs:-1.142397803413702, 1.2032049533611322) [,
color={rgb,1:red,0.00000000;green,0.00000000;blue,0.00000000}, draw opacity=1.0,
rotate=0.0,
font={\fontsize{14 pt}{18.2 pt}\selectfont}
] {$S_1$};
\node at (axis cs:3.096853094551612, 1.2032049533611322) [,
color={rgb,1:red,0.00000000;green,0.00000000;blue,0.00000000}, draw opacity=1.0,
rotate=0.0,
font={\fontsize{14 pt}{18.2 pt}\selectfont}
] {$S_2$};
\end{axis}

\end{tikzpicture}

%% file: images/wiener-hopf-R.tex
\begin{tikzpicture}[]
\begin{axis}[height = {50.8mm}, legend pos = {north east}, ylabel = {}, xmin = {0.8427259403604291}, xmax = {355.9876178389522}, ymax = {1.8245267662946156}, ymode = {log}, xlabel = {$P$}, unbounded coords=jump,scaled x ticks = false,xlabel style = {font = {\fontsize{11 pt}{14.3 pt}\selectfont}, color = {rgb,1:red,0.00000000;green,0.00000000;blue,0.00000000}, draw opacity = 1.0, rotate = 0.0},log basis x=10,xmajorgrids = true,xtick = {1.0,3.162277660168379,10.0,31.622776601683793,100.0,316.22776601683796},xticklabels = {$10^{0.0}$,$10^{0.5}$,$10^{1.0}$,$10^{1.5}$,$10^{2.0}$,$10^{2.5}$},xtick align = inside,xticklabel style = {font = {\fontsize{8 pt}{10.4 pt}\selectfont}, color = {rgb,1:red,0.00000000;green,0.00000000;blue,0.00000000}, draw opacity = 1.0, rotate = 0.0},x grid style = {color = {rgb,1:red,0.00000000;green,0.00000000;blue,0.00000000},
draw opacity = 0.1,
line width = 0.5,
solid},axis lines* = left,x axis line style = {color = {rgb,1:red,0.00000000;green,0.00000000;blue,0.00000000},
draw opacity = 1.0,
line width = 1,
solid},scaled y ticks = false,ylabel style = {font = {\fontsize{11 pt}{14.3 pt}\selectfont}, color = {rgb,1:red,0.00000000;green,0.00000000;blue,0.00000000}, draw opacity = 1.0, rotate = 0.0},log basis y=10,ymajorgrids = true,ytick = {0.001,0.01,0.1,1.0},yticklabels = {$10^{-3}$,$10^{-2}$,$10^{-1}$,$10^{0}$},ytick align = inside,yticklabel style = {font = {\fontsize{8 pt}{10.4 pt}\selectfont}, color = {rgb,1:red,0.00000000;green,0.00000000;blue,0.00000000}, draw opacity = 1.0, rotate = 0.0},y grid style = {color = {rgb,1:red,0.00000000;green,0.00000000;blue,0.00000000},
draw opacity = 0.1,
line width = 0.5,
solid},axis lines* = left,y axis line style = {color = {rgb,1:red,0.00000000;green,0.00000000;blue,0.00000000},
draw opacity = 1.0,
line width = 1,
solid},    xshift = 0.0mm,
    yshift = 0.0mm,
    axis background/.style={fill={rgb,1:red,1.00000000;green,1.00000000;blue,1.00000000}}
,legend style = {color = {rgb,1:red,0.00000000;green,0.00000000;blue,0.00000000},
draw opacity = 1.0,
line width = 1,
solid,fill = {rgb,1:red,1.00000000;green,1.00000000;blue,1.00000000},font = {\fontsize{8 pt}{10.4 pt}\selectfont}},colorbar style={title=}, xmode = {log}, ymin = {0.0003597602475616756}, width = {71.12mm}]\addplot+ [color = {rgb,1:red,0.00000000;green,0.60560316;blue,0.97868012},
draw opacity = 1.0,
line width = 2.0,
solid,mark = none,
mark size = 2.0,
mark options = {
    color = {rgb,1:red,0.00000000;green,0.00000000;blue,0.00000000}, draw opacity = 1.0,
    fill = {rgb,1:red,0.00000000;green,0.60560316;blue,0.97868012}, fill opacity = 1.0,
    line width = 1,
    rotate = 0,
    solid
}]coordinates {
(1.0, 1.4331375301072282)
(2.0, 0.06813782134881932)
(3.0, 0.05525628108559994)
(4.0, 0.03491524748213642)
(5.0, 0.029761229919360907)
(6.0, 0.023411058059259488)
(7.0, 0.02069268141189415)
(8.0, 0.017586183852559587)
(9.0, 0.015915154938420283)
(10.0, 0.014072662282210248)
(11.0, 0.012943720794518738)
(12.0, 0.01172456080524014)
(13.0, 0.010911568369302446)
(14.0, 0.010045434696328908)
(15.0, 0.009432401424016074)
(16.0, 0.008785494410749446)
(17.0, 0.008306905967421747)
(18.0, 0.007805427400020898)
(19.0, 0.0074215221409250276)
(20.0, 0.007021438221873089)
(21.0, 0.006706702233047368)
(22.0, 0.006380116320996422)
(23.0, 0.006117433980028548)
(24.0, 0.005845817164467348)
(25.0, 0.005623282116318876)
(26.0, 0.005393847065376855)
(27.0, 0.00520292250366918)
(28.0, 0.005006560282761399)
(29.0, 0.004840966830769559)
(30.0, 0.004671013510913903)
(31.0, 0.004526029962797048)
(32.0, 0.0043774981852841895)
(33.0, 0.0042495068232344)
(34.0, 0.004118589956277077)
(35.0, 0.004004772197666513)
(36.0, 0.003888514589553062)
(37.0, 0.003786642037467992)
(38.0, 0.0036827138999165135)
(39.0, 0.0035910016577582315)
(40.0, 0.0034975418000749332)
(41.0, 0.0034145434389031364)
(42.0, 0.0033300471403667)
(43.0, 0.0032545781499949585)
(44.0, 0.0031778157489690158)
(45.0, 0.0031088968546177658)
(46.0, 0.003038853668036453)
(47.0, 0.0029756682498190156)
(48.0, 0.0029114995770514844)
(49.0, 0.0028533612633538986)
(50.0, 0.0027943582344243808)
(51.0, 0.002740685942861368)
(52.0, 0.002686249280264022)
(53.0, 0.002636547783776672)
(54.0, 0.0025861674185354046)
(55.0, 0.002540012060971948)
(56.0, 0.0024932511340016767)
(57.0, 0.0024502756974002313)
(58.0, 0.0024067578838214087)
(59.0, 0.002366644874628959)
(60.0, 0.0023260442529214618)
(61.0, 0.002288517062771469)
(62.0, 0.0022505499521156035)
(63.0, 0.002215366484402171)
(64.0, 0.0021797848181828805)
(65.0, 0.002146732270430246)
(66.0, 0.0021133181789453207)
(67.0, 0.0020822087436839247)
(68.0, 0.0020507700962713645)
(69.0, 0.002021437397207948)
(70.0, 0.001991804111275105)
(71.0, 0.001964100232140881)
(72.0, 0.001936121199493499)
(73.0, 0.0019099141937099977)
(74.0, 0.001883454707019166)
(75.0, 0.0018586264998095624)
(76.0, 0.0018335660867962046)
(77.0, 0.0018100106994539764)
(78.0, 0.0017862412913828782)
(79.0, 0.0017638633314350576)
(80.0, 0.001741287707233889)
(81.0, 0.0017200010791894222)
(82.0, 0.0016985315379938295)
(83.0, 0.0016782583379822643)
(84.0, 0.001657815561919725)
(85.0, 0.0016384851263345733)
(86.0, 0.0016189972024894582)
(87.0, 0.0016005452861787217)
(88.0, 0.0015819468623374361)
(89.0, 0.001564314926219579)
(90.0, 0.0015465464795284215)
(91.0, 0.0015296810710096656)
(92.0, 0.0015126882723120762)
(93.0, 0.0014965404847014857)
(94.0, 0.0014802736442659735)
(95.0, 0.001464798643685217)
(96.0, 0.001449212226420257)
(97.0, 0.0014343688365775595)
(98.0, 0.0014194210367811963)
(99.0, 0.0014051713735205847)
(100.0, 0.0013908237408055346)
(101.0, 0.0013771328896055174)
(102.0, 0.0013633499989582785)
(103.0, 0.0013501857296070545)
(104.0, 0.001336934889623514)
(105.0, 0.0013242674031647326)
(106.0, 0.0013115183974069162)
(107.0, 0.0012993201011775525)
(108.0, 0.0012870449583467839)
(109.0, 0.0012752902655362834)
(110.0, 0.001263463054786678)
(111.0, 0.0012521282054537525)
(112.0, 0.0012407248536943887)
(113.0, 0.0012297877546101036)
(114.0, 0.001218785883089689)
(115.0, 0.0012082259641361245)
(116.0, 0.0011976047419764338)
(117.0, 0.0011874028271381552)
(118.0, 0.0011771428398008143)
(119.0, 0.0011672810310491733)
(120.0, 0.0011573641619899465)
(121.0, 0.001147825734580392)
(122.0, 0.0011382350585726678)
(123.0, 0.0011290043664688595)
(124.0, 0.0011197240532766538)
(125.0, 0.0011107864435762028)
(126.0, 0.0011018016708240139)
(127.0, 0.0010931434062022662)
(128.0, 0.0010844402804335758)
(129.0, 0.0010760484687422239)
(130.0, 0.0010676139537815846)
(131.0, 0.0010594764840446053)
(132.0, 0.0010512983358890557)
(133.0, 0.00104340382002709)
(134.0, 0.0010354705275811137)
(135.0, 0.0010278082472755096)
(136.0, 0.0010201089783282021)
(137.0, 0.0010126688365031588)
(138.0, 0.001005193388412978)
(139.0, 0.0009979658648749865)
(140.0, 0.0009907046194926378)
(141.0, 0.0009836807303166472)
(142.0, 0.0009766246127247095)
(143.0, 0.0009697958730237431)
(144.0, 0.0009629363137230527)
(145.0, 0.0009562947034764992)
(146.0, 0.0009496236036873914)
(147.0, 0.000943161536338776)
(148.0, 0.0009366712361219022)
(149.0, 0.0009303815296885145)
(150.0, 0.0009240647786215246)
(151.0, 0.0009179406290832031)
(152.0, 0.0009117905592577047)
(153.0, 0.0009058255160195945)
(154.0, 0.0008998356171462301)
(155.0, 0.0008940235603751663)
(156.0, 0.0008881876568083004)
(157.0, 0.0008825227765506998)
(158.0, 0.0008768350060618746)
(159.0, 0.000871311782809088)
(160.0, 0.0008657665769550192)
(161.0, 0.0008603797637636674)
(162.0, 0.0008549718297121412)
(163.0, 0.0008497164355759308)
(164.0, 0.0008444407392643967)
(165.0, 0.000839312013698575)
(166.0, 0.0008341637642120267)
(167.0, 0.0008291571829381004)
(168.0, 0.0008241318180054357)
(169.0, 0.0008192430696471109)
(170.0, 0.0008143362421634253)
(171.0, 0.0008095612158702629)
(172.0, 0.0008047687813611892)
(173.0, 0.0008001035552848658)
(174.0, 0.0007954215602370444)
(175.0, 0.0007908623907942827)
(176.0, 0.000786287061783567)
(177.0, 0.0007818303736420801)
(178.0, 0.0007773581071950206)
(179.0, 0.0007730004839271431)
(180.0, 0.0007686278370609438)
(181.0, 0.0007643660124131806)
(182.0, 0.0007600896937996833)
(183.0, 0.0007559205435318625)
(184.0, 0.0007517374052391215)
(185.0, 0.0007476579394898235)
(186.0, 0.0007435649692578493)
(187.0, 0.0007395723253971865)
(188.0, 0.0007355666394069536)
(189.0, 0.0007316580753402797)
(190.0, 0.0007277369114411395)
(191.0, 0.0007239097993318422)
(192.0, 0.0007200705106918609)
(193.0, 0.0007163223310719532)
(194.0, 0.0007125623802230359)
(195.0, 0.0007088907164641589)
(196.0, 0.0007052076697114807)
(197.0, 0.0007016102028314854)
(198.0, 0.0006980017250026334)
(199.0, 0.0006944762287832495)
(200.0, 0.0006909400782928799)
(201.0, 0.0006874844146878268)
(202.0, 0.0006840184388962739)
(203.0, 0.000680630553708727)
(204.0, 0.0006772326845541754)
(205.0, 0.0006739106033672132)
(206.0, 0.0006705788532519953)
(207.0, 0.0006673206775922872)
(208.0, 0.0006640531355081559)
(209.0, 0.0006608570392301351)
(210.0, 0.0006576518670956706)
(211.0, 0.0006545160929696744)
(212.0, 0.0006513715222094094)
(213.0, 0.0006482943787016456)
(214.0, 0.0006452087069678353)
(215.0, 0.000642188565198314)
(216.0, 0.0006391601533073226)
(217.0, 0.0006361954441762309)
(218.0, 0.0006332227132142788)
(219.0, 0.000630311924684679)
(220.0, 0.0006273933532402644)
(221.0, 0.0006245350277676009)
(222.0, 0.0006216691493597088)
(223.0, 0.0006188618814589284)
(224.0, 0.0006160472820733005)
(225.0, 0.0006132897160153341)
(226.0, 0.0006105250317714472)
(227.0, 0.000607815859395332)
(228.0, 0.0006050997743727981)
(229.0, 0.0006024377330402274)
(230.0, 0.0005997689771476078)
(231.0, 0.0005971528477466323)
(232.0, 0.0005945301947634553)
(233.0, 0.0005919587998648333)
(234.0, 0.0005893810655850495)
(235.0, 0.0005868532676878824)
(236.0, 0.0005843193080951161)
(237.0, 0.0005818340079373002)
(238.0, 0.0005793427175339753)
(239.0, 0.000576898852488662)
(240.0, 0.0005744491627018755)
(241.0, 0.0005720457052791705)
(242.0, 0.0005696365829192009)
(243.0, 0.0005672725393208595)
(244.0, 0.0005649029851158023)
(245.0, 0.0005625773938607747)
(246.0, 0.0005602464410881994)
(247.0, 0.000557958371724508)
(248.0, 0.00055566508491644)
(249.0, 0.0005534136367644659)
(250.0, 0.0005511571104107331)
(251.0, 0.0005489414113922869)
(252.0, 0.0005467207688001864)
(253.0, 0.0005445399742924717)
(254.0, 0.0005423543664069656)
(255.0, 0.0005402076581989639)
(256.0, 0.0005380562625459574)
(257.0, 0.000535942847800634)
(258.0, 0.000533824867443249)
(259.0, 0.0005317439777075163)
(260.0, 0.00052965864028007)
(261.0, 0.0005276095305858274)
(262.0, 0.0005255560873369517)
(263.0, 0.0005235380352800825)
(264.0, 0.0005215157601827133)
(265.0, 0.0005195280650929327)
(266.0, 0.0005175362539890744)
(267.0, 0.0005155782361077847)
(268.0, 0.0005136162058907206)
(269.0, 0.0005116872056164168)
(270.0, 0.0005097542934505643)
(271.0, 0.0005078536705751744)
(272.0, 0.0005059492331490885)
(273.0, 0.000504076366164609)
(274.0, 0.000502199778977869)
(275.0, 0.0005003540643810834)
(276.0, 0.0004985047210616382)
(277.0, 0.0004966855727306449)
(278.0, 0.0004948628843815657)
(279.0, 0.0004930697329429927)
(280.0, 0.0004912731275229657)
(281.0, 0.0004895054197459979)
(282.0, 0.00048773434145005717)
(283.0, 0.0004859915397015122)
(284.0, 0.00048424544841909126)
(285.0, 0.0004825270301025815)
(286.0, 0.0004808054008477743)
(287.0, 0.0004791108578853361)
(288.0, 0.00047741318028124663)
(289.0, 0.0004757420186129189)
(290.0, 0.0004740677963718967)
(291.0, 0.0004724195354786385)
(292.0, 0.00047076828592939023)
(293.0, 0.00046914245835876216)
(294.0, 0.0004675137119907876)
(295.0, 0.00046590986293631765)
(296.0, 0.0004643031629453424)
(297.0, 0.00046272084981176703)
(298.0, 0.00046113575167013265)
(299.0, 0.00045957454365610134)
(300.0, 0.00045801061470767795)
};
\addlegendentry{$\frac{|\mathfrak R^P - \mathfrak R |}{|\mathfrak R|}$}
\end{axis}

\end{tikzpicture}